\documentclass[times,review,nopreprintline,authoryear]{elsarticle}

\usepackage{jasr}
\usepackage{framed,multirow}

\usepackage{amssymb}
\usepackage{amsmath}
\usepackage{latexsym}
\usepackage{wasysym}
\usepackage{svg}
\usepackage[switch]{lineno}

\usepackage{url}
\usepackage{xcolor}
\usepackage{soul}
\definecolor{newcolor}{rgb}{.8,.349,.1}

\usepackage[citebordercolor=white]{hyperref}

\begin{document}

\verso{Edoardo Gramigna \textit{et al.}}

\begin{frontmatter}

\title{Analysis of NASA's DSN Venus Express radio occultation data for year 2014}

\author[1]{Edoardo \snm{Gramigna}\corref{cor1}}
\cortext[cor1]{Corresponding author: 
  Tel.: +39-331-281-6635;}
\ead{edoardo.gramigna@unibo.it}
\author[2]{Marzia \snm{Parisi}}
\ead{marzia.parisi@jpl.nasa.gov}
\author[2]{Dustin \snm{Buccino}}
\ead{dustin.r.buccino@jpl.nasa.gov}
\author[3]{Luis \snm{Gomez Casajus}}
\ead{luis.gomezcasajus@unibo.it}
\author[1,3]{Marco \snm{Zannoni}}
\ead{m.zannoni@unibo.it}
\author[4,5]{Adrien \snm{Bourgoin}}
\ead{adrien.bourgoin@obspm.fr}
\author[1,3]{Paolo \snm{Tortora}}
\ead{paolo.tortora@unibo.it}
\author[2]{Kamal \snm{Oudrhiri}}
\ead{kamal.oudrhiri@jpl.nasa.gov}

\address[1]{Department of Industrial Engineering, Alma Mater Studiorum - Università di Bologna, Via Fontanelle 40, I-47121 Forlì, Italy}
\address[2]{Jet Propulsion Laboratory, California Institute of Technology, Pasadena, California 91109}
\address[3]{Interdepartmental Center for Industrial Research in Aerospace, Alma Mater Studiorum - Università di Bologna, Via B. Carnaccini 12, I-47121 Forlì, Italy}
\address[4]{SYRTE, Observatoire de Paris, PSL Research University, CNRS, Sorbonne Universités, UPMC P6, LNE, 61 avenue de l’Observatoire, 75014 Paris, France}
\address[5]{Département d’Astrophysique-AIM, CEA/DRF/IRFU, CNRS/INSU, Université Paris-Saclay, Université de Paris, 91191 Gif-sur-Yvette, France}

\received{}
\finalform{}
\accepted{}
\availableonline{}
\communicated{}

\begin{abstract}
The Venus Express Radio Science Experiment (VeRa) was part of the scientific payload of the Venus Express (VEX) spacecraft and was targeted at the investigation of Venus' atmosphere, surface, and gravity field as well as the interplanetary medium. This paper describes the methods and the required calibrations applied to VEX-VeRa raw radio occultation data used to retrieve vertical profiles of Venus' ionosphere and neutral atmosphere. In this work we perform an independent analysis of a set of 25 VEX, single-frequency (X-band), occultations carried out in 2014, recorded in open-loop at the NASA Deep Space Network. Our temperature, pressure and electron density vertical profiles are in agreement with previous studies available in the literature. Furthermore, our analysis shows that Venus' ionosphere is more influenced by the day/night condition than the latitude variations, while the neutral atmosphere experiences the opposite. Our scientific interpretation of these results is based on two major responsible effects: Venus' high thermal inertia and the zonal winds. Their presence within Venus' neutral atmosphere determine why in these regions a latitude dependence is predominant on the day/night condition. On the contrary, at higher altitudes the two aforementioned effects are less important or null, and Venus' ionosphere shows higher electron density peaks in the probed day-time occultations, regardless of the latitude.
\end{abstract}


\end{frontmatter}


\section{Introduction}
\label{sec1}
Venus has been one of the prime targets in the early Solar System exploration, as the closest, and yet very different, planet. Between the 1960s and the 1980s intensive space mission campaigns have been carried out by the Soviet Union and the United States, which sent more than 30 spacecraft to study the so-called “Earth’s sister”. Venus revealed similarities in size, density, mass, volume, orbital radius and bulk composition to Earth. However, the similarities ended there. The missions discovered an atmosphere characterized by extremely high temperatures, pressures, and composition which renders it uninhabitable, pointing out two planets that had evolved very differently.

The first radio occultation experiment on Venus was performed in 1967 during a flyby of \textit{Mariner V} \citep{Fjeldbo1969,Academy1971,Barth1967,Kliore1967}, which discovered dayside and nightside ionization distributions in the upper atmosphere, as well as temperature and pressure profiles of the lower atmosphere of Venus. Subsequent missions such as \textit{Mariner X} \citep{Fjeldbo1975,Woo1975}, \textit{Venera} \citep{Yakovlev1991,Gavrik2009,Kolosov1979}, \textit{Pioneer Venus Orbiter} \citep{Kliore1979,Kliore1980,Newman1984} and \textit{Magellan} \citep{Hinson1995,1997DPS,Jenkins1994} contributed to new radio occultation investigations, increasing the understanding of the planet's atmosphere. In addition, the latter in the 1990s mapped the entire gravity field of the planet, up to the degree and order 180 \citep{Konopliv1999}. Since then, Venus remains unvisited for more than a decade as the priority for investigations of terrestrial planets shifted toward Mars, characterized by a more habitable environment. Nevertheless, there are still a large number of fundamental questions to be answered about the past, present and future of Venus. In 2005 the \textit{European Space Agency} (ESA) launched the \textit{Venus Express} mission to unveil the unsolved mysteries regarding the atmosphere, the plasma environment and the surface temperatures of Venus \citep{Titov2006,Svedhem2007}. One of the main objectives was studying the atmosphere, ionosphere and gravity of Venus through the \textit{VeRa} radio science instrument, by using S- and X-band (2.3 and 8.4 GHz, respectively) radio links between the spacecraft and Earth-based Deep Space Antennas \citep{Hausler2006}. 

The first results from the VeRa occultation experiments have been presented by \citet{Patzold2007}, who showed vertical profiles which revealed Venus' ionospheric structure between 100-500 km altitude and neutral atmosphere between 40-90 km altitude. Detailed studies have been carried out from the profiles retrieved in the following VeRa occultation seasons, which among other things show temperature on Venus which are latitude-dependent and a day/night variability of the electron density in the Venus' ionosphere \citep{Peter2014,Gerard2017}. \citet{Tellmann2009} showed also studies on the static stability of the atmosphere, which have been found to be latitude-dependent and nearly adiabatic in the middle cloud region. The static stability of the neutral atmosphere is also linked to the height and temperature of the tropopause, which is latitude-dependent too. Several studies from VeRa data have been conducted on the dynamics of the atmosphere of Venus, its thermal structure and its fluctuations due to gravity waves \citep{Piccialli2012,Tellmann2012,Lee2012,Ando2020}.

Nowadays, the exploration of Venus is still ongoing through the \textit{Akatsuki} mission from the \textit{Japanese Aerospace Exploration Agency} (JAXA), which reached the planet in 2015 \citep{Imamura2011,Imamura2017,Ando2020}. The near future foresees Venus as one of the main target for forthcoming space exploration missions. In particular, \textit{NASA Discovery Program} has selected two Venus space missions, namely \textit{VERITAS} from  \textit{NASA-Jet Propulsion Laboratory} and  \textit{DAVINCI+} from \textit{NASA Goddard Space Flight Center}, while the European Space Agency selected \textit{EnVision} Venus orbiter as the next Medium-class mission.

In this work we present 25 vertical profiles from VeRa occultation season of 2014, acquired by the \textit{NASA Deep Space Network} (DSN) managed by \textit{Jet Propulsion Laboratory} (JPL), in a collaboration between JPL and \textit{Alma Mater Studiorum - University of Bologna} (UNIBO). The retrieved vertical profiles are used to investigate the relations between the atmospheric parameters, the latitude and the day/night variations on Venus. For this investigation, we used one-way, single frequency (X-band) signals from VEX, recorded in open loop at the DSN stations. To our knowledge our work presents, for the first time singularly and independently, the 2014 VEX-DSN open-loop profiles. Moreover, our results come from an independent analysis, obtained from data received at a different ground station complex (NASA-DSN with respect to the ESA New Norcia (NNO), used as principal ground station in the VEX mission) and with non-identical signal reception/processing hardware and methods (our DSN open-loop vs ESA NNO closed-loop data). Section 2 reports the concept of radio occultation experiments, the adopted method, the developed algorithm, the data set, as well as the signal processing and the calibration of the observed frequencies. Section 3 presents the profiles obtained from VEX radio occultations in 2014, the results and our scientific interpretation. Section 4 presents our conclusions and discussions, while the Appendix shows our error analysis, which estimates the temperatures and pressure uncertainties of the retrieved profiles.





\section{Methods}
Atmospheric radio occultation investigations rely on the measurement of the frequency changes on a radio signal as it travels through the atmosphere of a planetary body. This technique is a powerful tool that has been used to probe remotely the planetary atmospheres since \textit{Mariner IV} in 1965 \citep{Kliore1965,Fjeldbo1966Mariner}.

\begin{figure}
	\centering
	\includegraphics[scale=0.2]{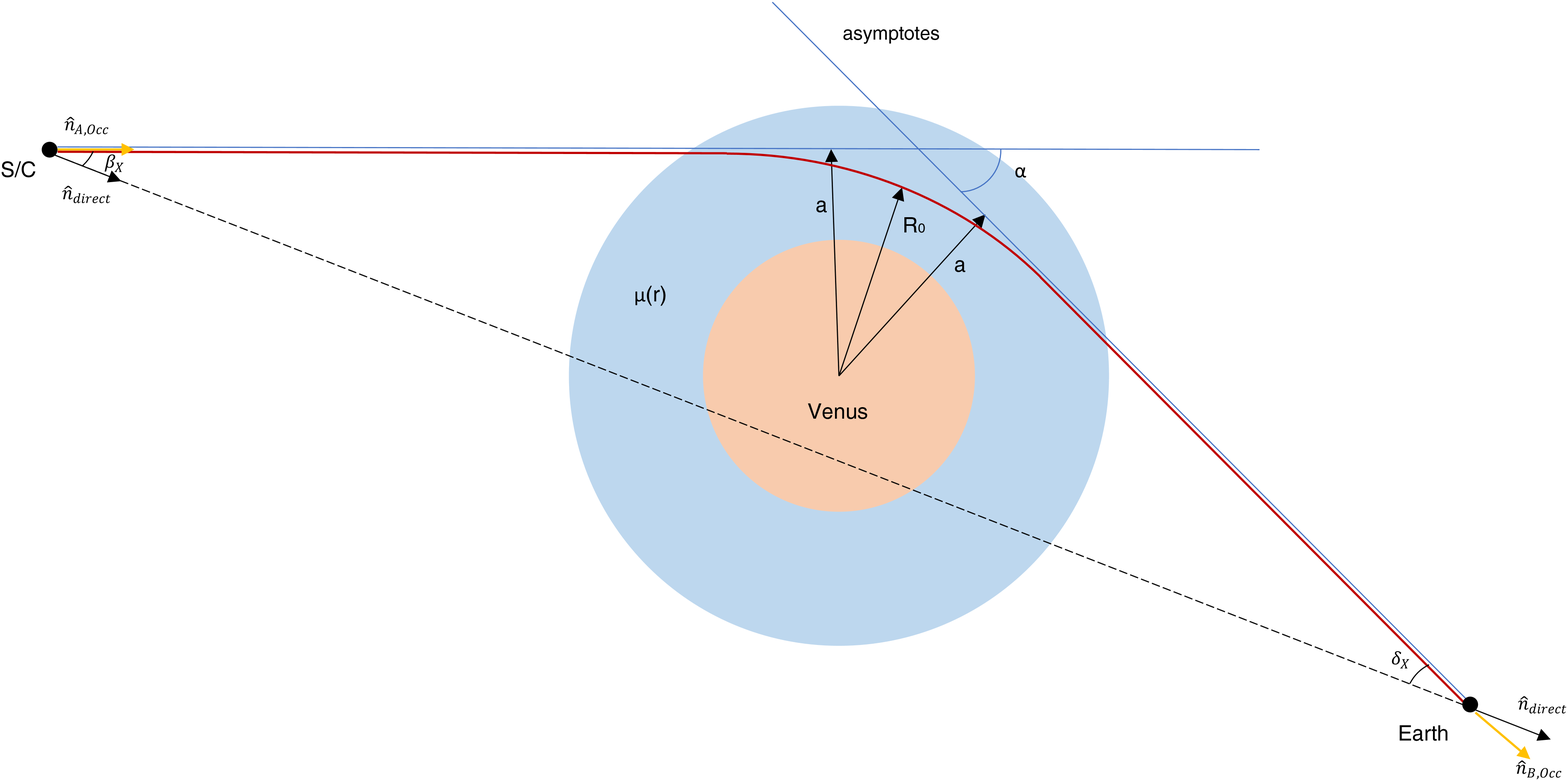}
	\caption{Ray bending in the atmosphere of Venus. $R_0$ is the ray path closest approach distance; $\alpha$ is the bending angle; $a$ is the impact parameter; $\mu$ is the index of refraction. The unit vectors showing the radio ray path would follow in vacuum are labelled as \textit{direct}, while the true path of the ray due to refraction at the target body is described by the unit vectors \textit{Occ}. $\beta_X$ and $\delta_X$ represents the two unknowns of the occultation experiment to be solved for each individual ray path, they are related to the geometry of the occultation and will yield the refractivity index of the target's atmosphere as a function of the impact parameter. The radii of the planet and the atmosphere are not to scale.}
	\label{fig1}
\end{figure}

The neutral atmosphere and ionospheric plasma around the planet have a refractive index different from 1, hence the radio frequency beam  experiences refraction. Refraction is mainly responsible for causing an excess path delay of the signal transmitted by a radio antenna (on board a spacecraft for a one-way down-link experiment). The change in the excess path delay during the occultation experiment generates a shift of the transmitted frequency that is interpreted as a bending of the light rays trajectory. The signal is eventually recorded by another antenna (an Earth-based ground station for a one-way down-link) and is then analyzed in order to extract the information about the crossed optical medium. The geometry of an occultation experiment is depicted in Figure \ref{fig1}.

The general methods and procedures to analyse and process radio occultation data, in order to determine vertical profiles of atmospheric properties from time series of frequency residuals have been presented by \citet{Phinney1968,Academy1971,Eshleman1973,Hausler2006,Withers2010,Withers2014,Dalba2019,Withers2020}. Next sections present details of our atmosphere retrieval algorithm, which includes all relativistic effects of the order of $1/c^2$, where $c$ is the speed of light in vacuum. The algorithm is developed in MATLAB environment. It first evaluates the radio occultation geometry from NASA’s SPICE toolkit Navigation and Ancillary Information Facility (NAIF) \citep{Acton1996}, namely the coordinate time of emission and reception, the relative positions and velocities of the transmitter, the receiver and the planets. These are used together with the frequency residuals to obtain the impact parameter $a$ and the bending angle $\alpha$ of the ray path during the occultation, see Figure \ref{fig1}. From the evolution of the bending as a function of the impact parameter, the target's atmosphere refractive index is eventually reconstructed using an Abel inversion method.

\subsection{Frequency residuals}

Radio occultation experiments are based on the accurate computation of the frequency residuals ($\Delta f$), which are defined as the difference between the observed (measured) frequency ($f_o$), and the expected (computed) frequency ($f_c$), namely 
\begin{equation}	
	\label{eq1}		
	\Delta f = f_{o}-f_{c}\mathrm{.}
\end{equation}

The frequency residuals carry crucial information on the planet's atmosphere and they represent the main input to the atmosphere retrieval algorithm, which aims at deriving the vertical profiles of the atmospheric parameters. The frequency residuals are computed from $f_c$, the predicted received frequency, which is determined from the modeling of the received frequency. Within the framework of general relativity, $f_B$, the frequency which is received by an observer $B$, is related to $f_A$, the frequency which is transmitted by an emitter $A$, according to the following expression \citep{Schinder2015}:
\begin{equation}
	\label{eq11}
	f_{B}\,(\mathbf{\hat{n}}_A,\mathbf{\hat{n}}_B) = f_{A}\,{\left(1-\mathbf{\hat{n}}_B \cdot {\mathbf{v}_B / c } \over 1-\mathbf{\hat{n}}_A \cdot {\mathbf{v}_A / c }\right)} {\sqrt{1-2U(\mathbf{r}_A) / c^2-(v_A / c)^2 } \over \sqrt{1-2U(\mathbf{r}_B)/ c^2-(v_B / c)^2}}\mathrm{,}
\end{equation}
where $v_A=\Vert\mathbf{v}_A\Vert$ and $v_B=\Vert\mathbf{v}_B\Vert$. $\mathbf{r}_A$ and $\mathbf{v}_A$ are  the position and velocity vectors of the emitter with respect to the occulting body's center-of-mass, respectively. $\mathbf{r}_B$ and $\mathbf{v}_B$ are the position and velocity vectors of the receiver with respect to the occulting body's center-of-mass, respectively. The position and velocity of the emitter are given at time of emission $t_A$, namely $\mathbf{r}_A=\mathbf{r}_A(t_A)$ and $\mathbf{v}_A=\mathbf{v}_A(t_A)$, respectively, while the position and velocity of the receiver are given at time of reception $t_B$, namely $\mathbf{r}_A=\mathbf{r}_A(t_B)$ and $\mathbf{v}_A=\mathbf{v}_A(t_B)$, respectively. The unit-vectors $\mathbf{\hat{n}}_A$ and $\mathbf{\hat{n}}_B$ represent the direction of the radio ray at the level of emitter and receiver, respectively; they are both expressed in the occulting body's center-of-mass frame. $U(\mathbf{r})$ is the opposite of the total Newtonian gravitational potential evaluated at the position $\mathbf{r}$. The expression \ref{eq11} is valid for a stationary spacetime and include all terms at first-order in $1/c^2$. In the case of a down-link one-way radio occultation experiment at Venus, the emitter $A$ is a spacecraft, the receiver $B$ is the Earth-based ground station, the occulting body is Venus (see Figure \ref{fig1}), and $U(\mathbf{r})$ contains the gravitational potentials of the Earth, the Sun, Venus, Jupiter, Saturn and Mars.

Let us emphasize that the positions and velocities of planets, satellites, and spacecraft are usually integrated and then distributed in the solar system barycentric frame (see e.g., \cite{2021AJ....161..105P} for a presentation of DE440/441 planetary ephemerides). Therefore, in order to compute Eq. \ref{eq11} in Venus' center-of-mass frame, we apply appropriate Lorentz transformations on barycentric positions, velocities, and directions of the radio rays. These corrections starts with terms $\propto 1/c$ and must be taken into account to ensure that Eq. \ref{eq11} is coherent up to first-order in $1/c^2$.

Hereafter, we use $T$ to denote the barycentric coordinate time and $t$ to refer to the proper time at Venus' center of mass. We use the convention that a vector (e.g., a velocity or a direction tangent to the radio ray, etc.) expressed in Venus' center-of-mass frame is denoted by a ``\emph{bold lowercase letter}'' (e.g., $\mathbf{v}$ and $\mathbf{\hat{n}}$, for previous examples), while in the barycentric frame the same vector is denoted by a ``\emph{bold capital letter}'' (e.g., $\mathbf{V}$ and $\mathbf{\hat{N}}$, again for previous examples). We denote a position vector expressed in Venus' center-of-mass frame by $\mathbf{r}$, while the barycentric position vector pointing to the same point is denoted by $\mathbf{R}$. For instance, the barycentric position, velocity, and direction of the radio ray at the emission point $A$ are denoted by $\mathbf{R}_A$, $\mathbf{V}_A$, and $\mathbf{\hat{N}}_A$, respectively. We denote by $\mathbf{R}_{\venus}=\mathbf{R}_{\venus}(T_0)$ and $\mathbf{V}_{\venus}=\mathbf{V}_{\venus}(T_0)$ the barycentric positions and velocity vectors of Venus evaluated at the initial coordinate time $T_0$ for the Lorentz transformations. The magnitude of the barycentric velocity of Venus is denoted by $V_{\venus}=\Vert\mathbf{V}_{\venus}\Vert$. The relationships between quantities $t$, $\mathbf{r}$, $\mathbf{v}$, $\mathbf{\hat{n}}$ expressed in Venus' center-of-frame and $T$, $\mathbf{R}$, $\mathbf{V}$, $\mathbf{\hat{N}}$ expressed in the barycentric frame are given by
\begin{subequations}\label{eq:Lorentz}
\begin{align}
  t&=\Gamma\,\left[T-T_0-\frac{(\mathbf{R}-\mathbf{R}_{\venus})\cdot\mathbf{V}_{\venus}}{c^2}\right]\text{,}\\
  \mathbf{r}&=\textbf{R}-\mathbf{R}_{\venus}-\Gamma\,\mathbf{V}_{\venus}(T-T_0) +\left(\Gamma-1\right)\frac{[\mathbf{V}_{\venus}\cdot(\mathbf{R}-\mathbf{R}_{\venus})]\mathbf{V}_{\venus}}{V_{\venus}{}^2}\text{,}\\
  \mathbf{v}&=\Gamma^{-1}\left(1-\cfrac{\textbf{V}\cdot\mathbf{V}_{\venus}}{c^2}\right)^{-1}\left[\mathbf{V}-\Gamma\,\textbf{V}_{\venus}+\left(\Gamma-1\right)\frac{(\mathbf{V}_{\venus}\cdot\mathbf{V})\mathbf{V}_{\venus}}{V_{\venus}{}^2}\right]\text{,}\\
  \mathbf{\hat{n}}&=\Gamma^{-1}\left(1-\cfrac{\mathbf{\hat{N}}\cdot\mathbf{V}_{\venus}}{c}\right)^{-1}\left[\mathbf{\hat{N}}-\frac{\Gamma\,\mathbf{V}_{\venus}}{c}+\left(\Gamma-1\right)\frac{(\mathbf{V}_{\venus}\cdot\mathbf{\hat{N}})\mathbf{V}_{\venus}}{V_{\venus}{}^2}\right]\text{,}
\end{align}
\end{subequations}
where $\Gamma$ is the Lorentz factor being defined such as $\Gamma=[1-(V_{\venus}/c)^2]^{-1/2}$. We call $\mathbf{V}_{\venus}$ the parameter of the Lorentz transformation.

Inverse Lorentz transformations to go from the Venus's center-of-mass frame to the barycentric frame are straightforwardly obtained by switching ``\emph{capital}'' quantities by their ``\emph{lowercase}'' counterparts and by reversing the sign of the barycentric velocity of Venus. For instance, the coordinate time at reception is given by $T_B=T_0+\Gamma\,(t_B+\mathbf{r}_B\cdot\mathbf{V}_{\venus}/c^2)$.

While using the transformations \ref{eq:Lorentz}, let us emphasize that we make the approximation that the barycentric motion of Venus is inertial during the time interval $(T_B-T_0)$. Therefore, to keep $(T_B-T_0)$ small, we can choose $T_A$, the date of emission of the signal, as the origin of the coordinate time for the Lorentz transformation, namely $T_0\equiv T_A$. To determine $T_A$ we proceed as follows. The reception time which is dated in \emph{Universal Coordinate Time} (UTC) can directly be transformed into a barycentric coordinate time (namely $T_B$) using planetary ephemerides. Therefore, $T_B$ is known from the data and DE440/441. Then, the coordinate time at emission is determined by solving iteratively the following light-time equation
\begin{equation}
  \label{eq:lighttime}
  T_A=T_B-\frac{\Vert\mathbf{R}_B(T_B)-\mathbf{R}_A(T_A)\Vert}{c}\text{,}
\end{equation}
assuming $T_A=T_B$ at the first iteration. In Eq. \ref{eq:lighttime}, we do not consider relativistic corrections from the Shapiro time delay which are $\propto 1/c^3$. We also omit the atmospheric delay which could nevertheless be accounted for as described in Appendix A of \cite{2021A&A...648A..46B}. Then, from $T_A$, we can now determine the parameters of the Lorentz transformations, namely $\mathbf{R}_{\venus}=\mathbf{R}_{\venus}(T_A)$ and $\mathbf{V}_{\venus}=\mathbf{V}_{\venus}(T_A)$. It is now straightforward to apply Eqs. \ref{eq:Lorentz} to transform all the barycentric positions and velocities of the problem into positions and velocities expressed in Venus' center-of-mass frame. After substituting these into Eq. \ref{eq11}, we get the expression of the received frequency $f_B$ which we might see as a function of $\mathbf{\hat{n}}_A$ and $\mathbf{\hat{n}}_B$ only.

The data processing can be divided into two parts: (i) the calibration of the observations and (ii) the inversion of the data. Both parts are based on two different realizations of the frequency residuals \ref{eq1}. In the first part, the computed frequency is determined assuming a radio signal propagating in vacuum, namely $\mathbf{\hat{n}}_A=\mathbf{\hat{n}}_B=\mathbf{\hat{n}}_{direct}$. In this picture, we define the computed frequency $(f_{c,direct})$ by the following expression:
\begin{equation}
  \label{eq:fc,direct}
  f_{c,direct}\equiv f_{B}\,(\mathbf{\hat{n}}_{direct},\mathbf{\hat{n}}_{direct})\text{,}
\end{equation}
where $\mathbf{\hat{n}}_{direct}$ is the unit-vector being tangent to the straight-line segment joining $A$ and $B$ (see figure \ref{fig1}). After substituting for $f_c$ from Eq. \ref{eq:fc,direct} into \ref{eq1}, we eventually derive the frequency residuals $\Delta f_{direct}$ which are used for calibrating the data. This point is further discussed in Sec. \ref{sec2.3}.


Once the data have been properly calibrated, we focus on the second part of the data processing, namely the inversion, which is based on the accurate computation of the received frequency, too. The main difference with respect to the calibration part relies on the fact that the radio signal does not propagate in vacuum anymore. As a matter of fact, the refraction effect is taken into account through the bending of the radio ray trajectory, that is to say $\mathbf{\hat{n}}_{A,Occ}\neq\mathbf{\hat{n}}_{B,Occ}$ (see Figure \ref{fig1}). In this picture, we define the computed frequency $(f_{c,Occ})$ by
\begin{equation}
  \label{eq:fc,Occ}
  f_{c,Occ}\equiv f_{B}\,(\mathbf{\hat{n}}_{A,Occ},\mathbf{\hat{n}}_{B,Occ})\text{.}
\end{equation}
In order to emphasize the atmospheric contribution in the frequency computation, it is convenient to re-express the received frequency $f_B$ in terms of $f_{c,direct}$, and in term of the deviation of the radio ray with respect to its vacuum trajectory, such as
\begin{equation}
  f_{B}\,(\mathbf{\hat{n}}_{A,Occ},\mathbf{\hat{n}}_{B,Occ})=f_{c,direct}\,\left(1-\cfrac{(\mathbf{\hat{n}}_{B,Occ}-\mathbf{\hat{n}}_{B,direct})\cdot\mathbf{v}_B/c}{1-\mathbf{\hat{n}}_{B,direct}\cdot\mathbf{v}_B/c}\right)\left(1-\cfrac{(\mathbf{\hat{n}}_{A,Occ}-\mathbf{\hat{n}}_{A,direct})\cdot\mathbf{v}_A/c}{1-\mathbf{\hat{n}}_{A,direct}\cdot\mathbf{v}_A/c}\right)^{-1}\text{.}
\end{equation}
After substituting for $f_c$ from \ref{eq:fc,Occ} into \ref{eq1}, we compute the frequency residuals $\Delta f_{Occ}$ for the occultation. Using a Newton-Raphson algorithm, we determine $\mathbf{\hat{n}}_{A,Occ}$ and $\mathbf{\hat{n}}_{B,Occ}$ by minimizing the residuals, that is to say $\Delta f_{Occ}=0$. This allows us to eventually derive the bending angle of the radio path according to
\begin{equation}
  \alpha=\arccos\left(\mathbf{\hat{n}}_{A,Occ}\cdot\mathbf{\hat{n}}_{B,Occ}\right)\text{.}
\end{equation}
From the bending angle and the impact parameter of the radio ray trajectory (whose expression is easily deduced from the pointing and positions vectors in Venus' centre-of-mass frame), the inversion proceeds with the Abel transform. This is the subject of the next section.


\subsection{Abel Inversion method}
The bending angle $\alpha$ and the impact parameter $a$ of each ray path represent the fundamental parameters of the occultation measurement. They are transformed in vertical profiles of the planet’s atmosphere refractive index, $\mu$, at the ray closest approach distance, through an Abel transform integral inversion formula \citep{Academy1971}, namely 

\begin{equation}
	\label{eq2}
	\pi \ln \mu(R_0) = \int_{a=a_0}^{a=\infty} \ln \left\{{a \over a_0} + \left[\left({a \over a_0}\right)^2 -1 \right ]^{1/2} \right \}\, {\mathrm{d}\alpha \over \mathrm{d}a} \,\mathrm{d}a\mathrm{,}
\end{equation}
where $R_0$ is the closest approach distance, which is computed through the Bouguer's rule \citep{Born1959,Academy1971, Kursinski2000}, namely
\begin{equation}
	\label{eqB}
	R_0 = {a_0 \over \mu(R_0)}\mathrm{.}
\end{equation}
Hence, a vertical profile $\mu(R)$ can be determined for any $R=R_0$ using the derived values of $\alpha$ and $a$.

The vertical profiles of the atmospheric parameters as refractivity, mass density, neutral density, temperature and pressure are easily and directly obtained directly from the refractive index, by making use of the hydrostatic equilibrium assumption together with the ideal gas law. Regarding the electron density, we assumed a linear relationship between the refractivity, $\nu_e$, and the electron density, $N_e$, see Section \ref{sec3}.

The main facts, assumptions and limitations of this method are summarized hereafter:
\begin{itemize}
	\item One-Way Link: The radio signal is transmitted from the spacecraft and received at the ground station;
	\item Single Frequency: only X-band signals are used in these experiments since they were performed using the Venus Express High Gain Antenna 2. This particular antenna did not support the S-band link, so only X-band data is available for the occultations we present in this work. This is a limitation for the ionosphere investigation, since a dual-frequency analysis provides always more reliable results. On the other hand, a single-frequency analysis is not a limitation for the neutral atmosphere since it is a non-dispersive medium;
	\item Stable Frequency Source: The source of the transmitted frequency must be stable (use of Ultra Stable Oscillators by the transmitter is key to the success of One-Way link occultations); 
	\item Spherically Symmetric atmosphere and ionosphere at the target: This assumption allows to assume that the radio signal will travel within the \textit{r-z} plane defined in Fig.1 of \citet{Withers2014}. If the assumption is not satisfied (as for the oblate planets Jupiter, Saturn, Uranus and Neptune), gradients of refractivity will exists perpendicular to the r-z plane and the radio signal will travel outside it. As a result, the adopted method would not be valid for those instances. Nevertheless, in the context of occultation analysis, this assumption is adequate for many atmospheres and ionospheres in the solar system, including those of Venus, the Earth, Mars, the four Galilean satellites, Titan, Enceladus, and Triton; 
	\item Well-mixed Atmosphere: Assumption used to convert the observed refractivity to the atmospheric density; 
	\item Ideal Gas Behavior: The neutral atmosphere is assumed to behave as an ideal gas.
\end{itemize}

\begin{figure}
	\centering
	\includegraphics[scale=0.6]{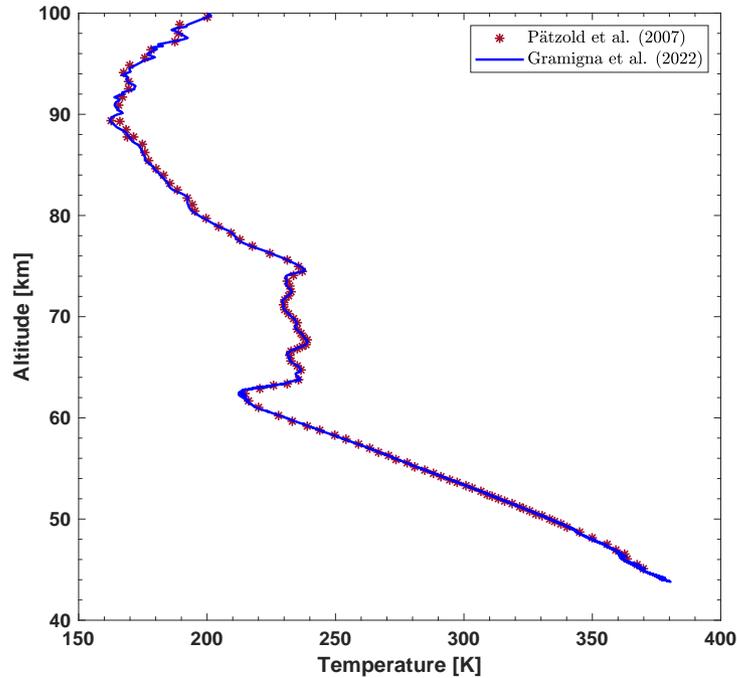}
	\caption{Software validation. The data correspond to a radio occultation of VEX on August 22, 2006. The blue curve is obtained from our own Abel inversion software, while the purple asterisks represent the temperature profile published in \citet{Patzold2007}. The authors acknowledge the Principal Investigator B. Haeusler of the VRA instrument onboard the Venus Express mission for providing datasets in the archive. Dataset of the VRA instrument has been downloaded from the ESA Planetary Science Archive \href{http://archives.esac.esa.int/psa}{http://archives.esac.esa.int/psa} \citep{BESSE2018131}.}
	\label{figcomparison}
\end{figure}

Our Abel inversion software has been tested and validated with respect to several radio occultation profiles presented in the literature. Figure \ref{figcomparison} shows a validation with respect to the VEX results published in \citet{Patzold2007}.

\subsection{Data set}
\begin{figure}
	\centering
	\includegraphics[scale=0.5]{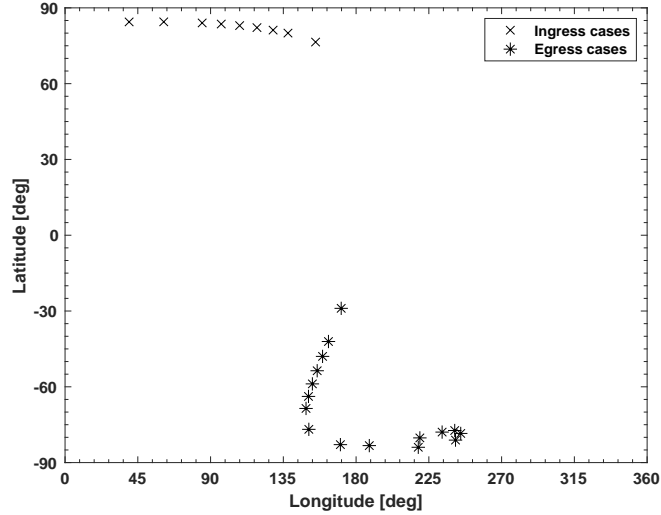}
	\caption{Spatial distribution of VEX 2014 occultations data as a function of latitude and longitude. Points are related to the occultation point at 50 km altitude.}
	\label{fig4}
\end{figure}

Our data-set comprises 25 occultations from the VeRa Venus Express radio science instrument, acquired between January and March 2014 occultation season. The data have been recorded in open-loop by the \textit{NASA Deep Space Network}, in particular by the Deep Space Stations (DSS) 34 and 43 from the Canberra complex in Australia, all during local night-time. It is worth to mention, however, that our data-set is not currently publicly available. The occultations are composed of 9 ingress points and 16 egress points. The former are all constrained in the north polar region, while the latter are characterized by a wide latitudinal coverage in the Southern hemisphere, see Figure \ref{fig4}. The data were recorded by the Radio Science Receiver (RSR), a computer-controlled open-loop receiver that digitally recorded the VEX signal with a sampling rate of 2 kHz, by means of an analog to digital converter (ADC) and multiple digital filter sub-channels. The RSR relies on frequency predicts to remain tuned to the incoming signal, so it does not have a feedback loop with which to track and lock the received signal, as for the closed-loop ones.  This is especially advantageous in the occultations investigations, where is challenging to establish and/or maintain the signal lock.  Furthermore, in order to retrieve the frequency time series, the RSR data were processed through a spectral FFT algorithm (Fast Fourier Transform) \citep{Paik2011} on consecutive intervals of 500 IQ samples, in order to obtain a sufficiently high number of frequency measurements. This resulted in a count-time, or integration time-step, of the frequency time-series of 0.25 seconds. 

Furthermore, the open-loop data allow to carefully address the potential presence of the multipath phenomenon at  Venus' tropopause, first discussed in \citet{imamura2018fine}. In particular, we carried out a signal processing analysis in the frequency domain, at different integration times. Our conclusions are  that our data set does not show multiple peaks (comparable in terms of power), also called subsignals, at the tropopause. Multiple peaks would have been a clear indication of the multipath phenomenon, and their absence led us to the conclusions that (a) multipath is not affecting our results to a level where it would be significant (i.e. its effect falls below our uncertainty level), and (b) a Full Spectrum Inversion (FSI) method \citep{imamura2018fine} is not required within this investigation.

\subsection{Calibrations}
\label{sec2.3}
The accuracy of a one-way radio occultation experiment relies on the stability of the spacecraft's transmitted frequency. Venus Express was equipped with an Ultra Stable Oscillator (USO), an instrument which guarantees a highly precise control of the transmitted radio frequency in terms of its Allan Standard Deviation over a wide range of integration times \citep{Hausler2006}. However, several noise sources and errors affect the observed frequency (i.e. thermal noise, media propagation noise, Earth's ionosphere and troposphere, spacecraft clock and estimated trajectory), which result in fluctuations in the evaluated frequency residuals. As a consequence, if not calibrated, these effects can jeopardize the accuracy of the retrieved atmospheric profiles. The calibration process is crucial to compensate the observed frequency for the noises and errors present in the signal, providing dependable results.

\begin{figure}
	\centering
	\includegraphics[scale=0.5]{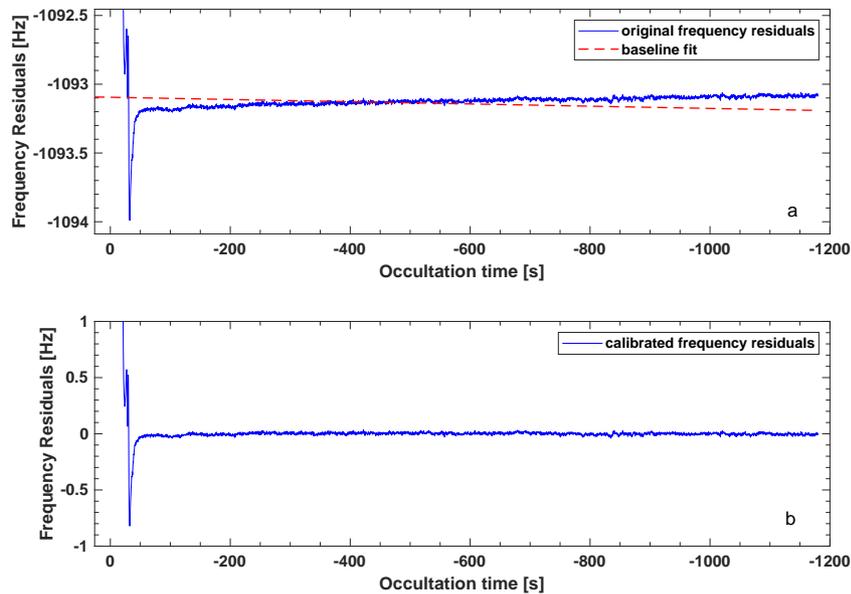}
	\caption{Baseline fit on the VEX DOY 032 Egress occultation frequency residuals. Panel \textbf{a}): zoom on the original frequency residuals, in which a bias and a linear trend is clearly visible. In red the baseline fit is depicted: we performed a first order polynomial and we applied it to the entire frequency residuals time series, whose coefficient are evaluated outside the atmosphere between -500 s and -1200 s. Panel \textbf{b}): calibrated frequency residuals after subtracting the baseline fit polynomial. They are characterized by a flat zero-mean baseline, as required. It is worth mentioning that the bias magnitude (visible before the calibration in Panel \textbf{a}):) is not relevant as long as it is calibrated with a proper baseline fit.}
	\label{fig44}
\end{figure}

The first calibration is performed to correct the frequency residuals for the local Earth's ionosphere and troposphere, which cause a delay in the spacecraft signal propagation. This delay is translated into a phase delay of the recorded signal and then converted into a frequency shift in the Doppler observables. These corrections are in the form of polynomials, whose coefficients are provided by the \textit{Tracking System Analytic Calibration (TSAC) Group} at JPL, based on Global Positioning Systems (GPS) data analysis. The remaining noises can be evaluated from the baseline of the frequency residuals, the region where the signal is traveling outside the atmosphere of the planet. In an ideal occultation experiment the baseline should be flat with low-noise, zero-mean frequency residuals. A zero-mean value indicates that the frequency residuals are fully calibrated for all non-atmospheric sources so that only the signature of the atmospheric refraction would remain. For this reason, a second calibration is required to compensate the spacecraft clock, the estimated trajectory, the plasma noise and thermal noise, with the goal to better estimate the transmitted frequency from the unperturbed signal travelling outside the planet's atmosphere. This second calibration is usually called \textit{baseline fit}: the polynomial coefficients of order \textit{n} are calculated from the baseline (the region outside the atmosphere), which are used to generate a new calibration polynomial for the entire observation time-span, which is then subtracted from the entire frequency residuals time series. The adopted polynomial's order for the baseline fit is between 0 and 2, depending on the quality of the frequency residuals. Figure \ref{fig44} shows the frequency residuals before and after the baseline fit, while Equation \ref{eq3} summarizes the performed calibration process:

\begin{equation}
	\label{eq3}
	\Delta f_{calibrated} = \Delta f + \Delta f_{tropo} - \Delta f_{iono} - \left[p_0 + p_1 t + p_2 t^2 + ... + p_n t^n \right]\mathrm{,}
\end{equation}
where $t$ is the time at the receiver related to the frequency residuals [$s$], $\Delta f_{tropo}$ and $\Delta f_{iono}$ are the troposphere and ionosphere calibration polynomials [Hz], while the polynomial of order $n$ represents the baseline fit, it is characterized by the polynomial coefficients $p_n$ [Hz $\cdot$ $s^{-n}$], and it compensates spacecraft clock and estimated trajectory errors, in addition to thermal and plasma noise.

To conclude, let us emphasize that even though the baseline fit is a subjective action, the profiles converge regardless of the adopted baseline fit, as long as a proper low-order (between order 0-2) calibration is performed.

\section{Data analysis results}
\label{sec3}
The first parameter obtained from the Abel transform, as a function of the altitude, is the refractive index $\mu$, which is used to evaluate the refractivity of the atmosphere:

\begin{equation}
	\nu(h) = \mu(h) -1\mathrm{.}
\end{equation}

The total refractivity $\nu$ is the sum of the refractivity of the ionosphere $\nu_e$ and the refractivity of the neutral atmosphere $\nu_n$:

\begin{equation}
	\nu = \nu_e + \nu_n\mathrm{.}
\end{equation}

\begin{figure}
	\centering
	\includegraphics[scale=0.5]{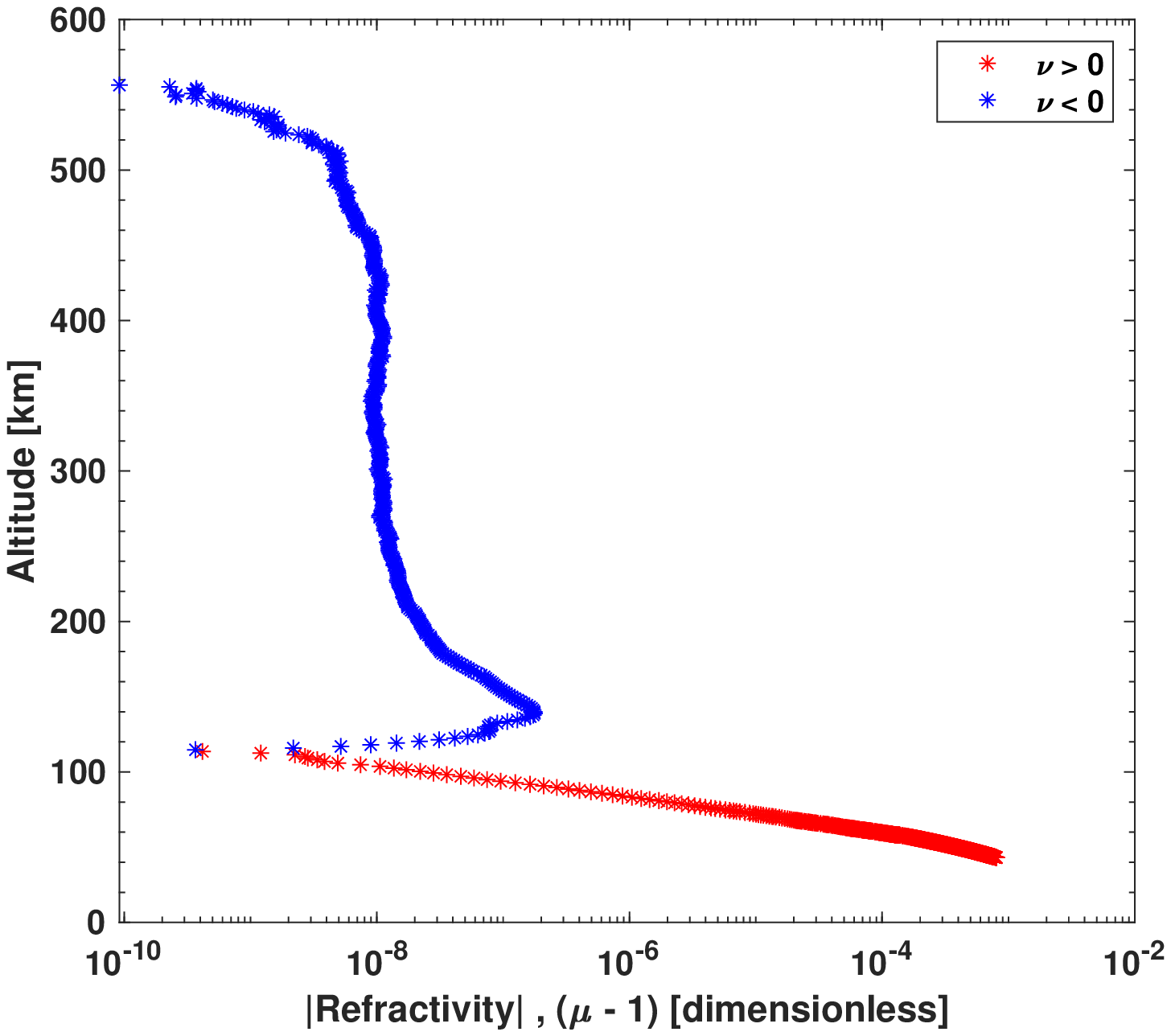} 
	\includegraphics[scale=0.5]{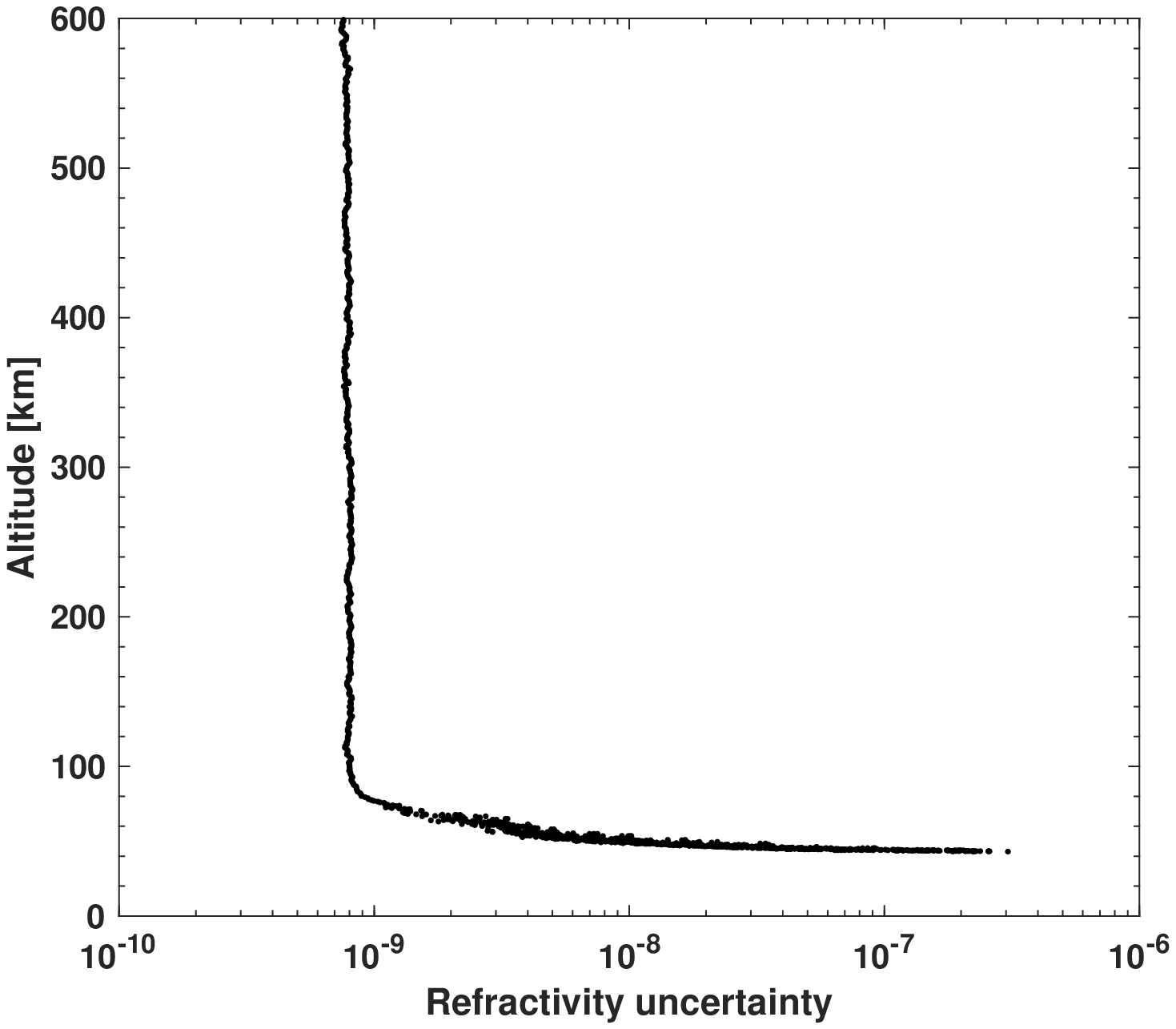}
	\centering
	\caption{Absolute value of the refractivity profile of VEX DOY 032 2014 egress occultation and related uncertainties. $\nu$ is negative in the ionosphere and positive in the neutral atmosphere. The altitude is evaluated with respect to the mean Venus radius of 6051.8 km. The uncertainties are retrieved from a Monte Carlo method, taking into account the noise present in the baseline of the frequency residuals, as well as a variable thermal noise function of altitude, see Appendix.}
	\label{fig5}
\end{figure}

The single-frequency occultation experiment assumes that if the derived value of $\nu$ is negative, then $\nu_e$ is assumed to be identical to the measured value of $\nu$ while $\nu_n$ is assumed to be zero and vice versa. In other words, we assume that the ionosphere and neutral atmosphere of Venus are characterized by a clear separation occurring at a certain altitude. In the intermediate region where $\nu$ is experimentally indistinguishable from zero both $\nu_e$ and $\nu_n$ are assumed to be zero \citep{Withers2010}.
A refractivity profile from the VEX egress occultation occurred on 1$^{st}$ February 2014 is shown in Figure \ref{fig5}, together with the related uncertainties. The uncertainties are obtained from a Monte Carlo simulation \citep{Schinder2011,Schinder2012}, a method which allows to propagate the uncertainties in the frequency shift through the data processing pipeline. We performed 2000 runs of a representative NASA-DSN 2014 VEX occultation, by adding gaussian random noise time-series to the original frequency residuals, in order to obtain the standard deviations of the profiles in terms of relevant atmospheric parameters. This error analysis considers the noise present in the baseline of the frequency residuals, namely including the USO noise, plasma noise, and a variable thermal noise as a function of altitude. The thermal noise evolves as the reciprocal of the signal-to-noise ratio: since the latter decreases across the atmospheric profile, the thermal noise is not constant in general, and it increases as the rays probe lower regions of the target's atmosphere. The details of the error analysis are presented in the Appendix.

\subsection{Ionosphere}
The ionosphere is a thin plasma structure within the extended neutral atmosphere and it is sensitive to radio waves above 100 km. The interaction of the electrons with the radio wave leads to an advance of the phase of the carrier frequency. The electron density $N_e$ in the ionosphere is retrieved from the refractivity of the ionosphere $\nu_e$:

\begin{equation}
	\nu_e(h) = { N_e(h) e^2 \over 8 \pi^2 m_e \epsilon_0 f^2}\mathrm{,}
\end{equation}
where $e$ is the elementary charge, $m_e$ is the electron mass,  $\epsilon_0$ is the permittivity of free space and $f$ is the transmitted frequency \citep{Withers2010}.

\begin{figure}
	\centering
	\includegraphics[scale=0.5]{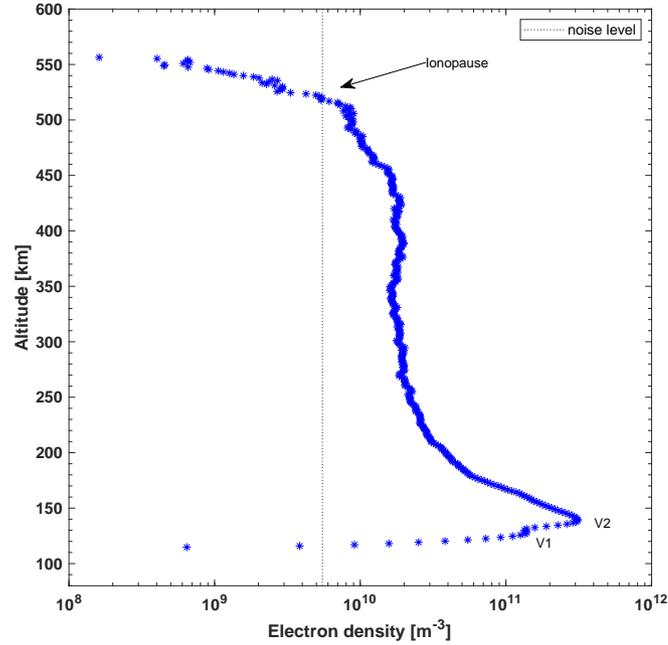}
	\caption{Electron density profile of VEX DOY 032 2014 egress occultation. Latitude (at the 50 km altitude occultation point) = -77.300$^{\circ}$; Longitude = 240.998$^{\circ}$; Solar Zenith Angle (SZA) = 74.5 deg. The mean noise level is $5.5 \cdot 10^9 el/m^3$ and it is evaluated from the baseline of the electron density profile.}
	\label{fig6}
\end{figure}

\begin{figure}
	\centering
	\includegraphics[scale=0.5]{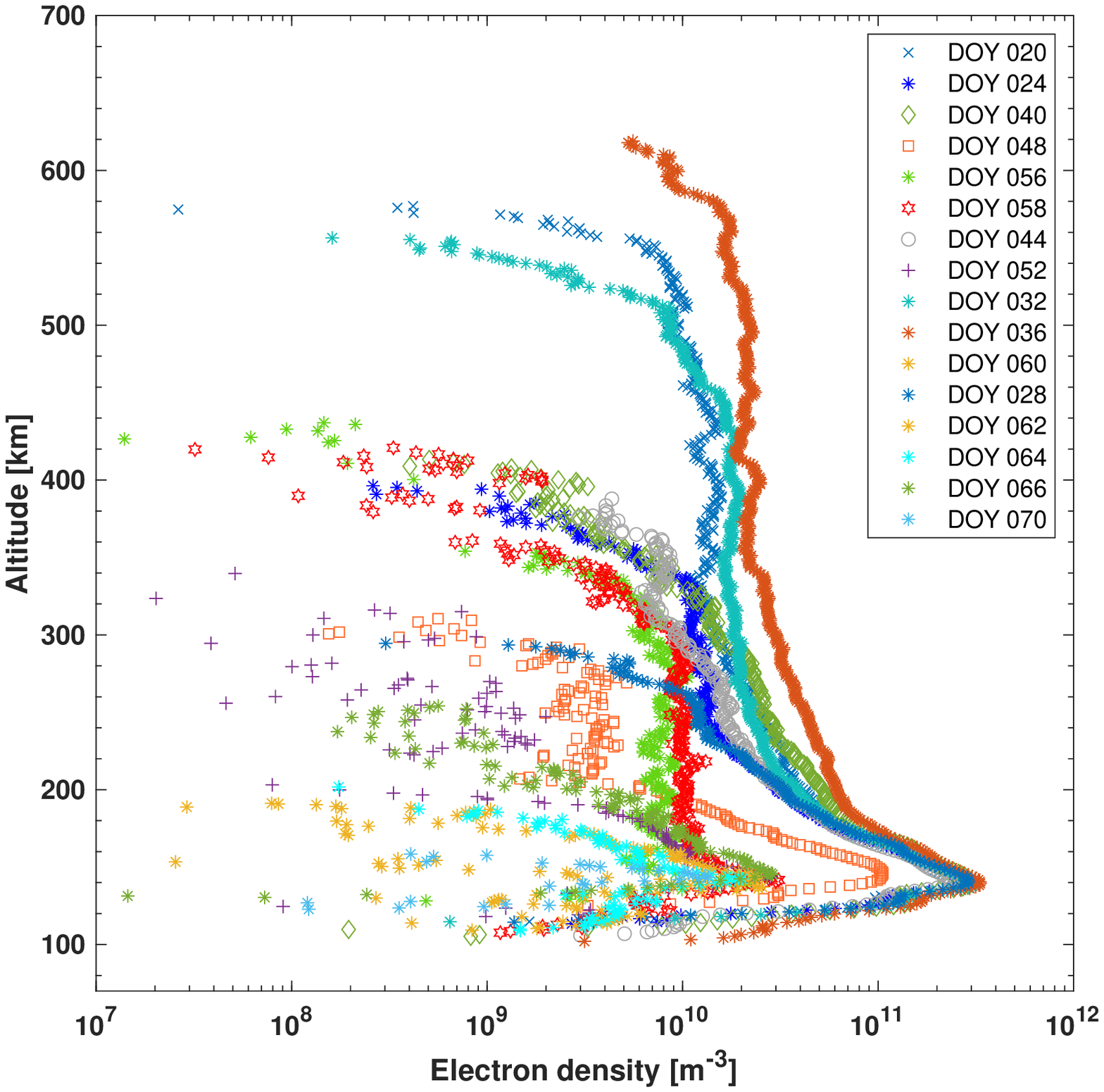}
	\centering
	\includegraphics[scale=0.5]{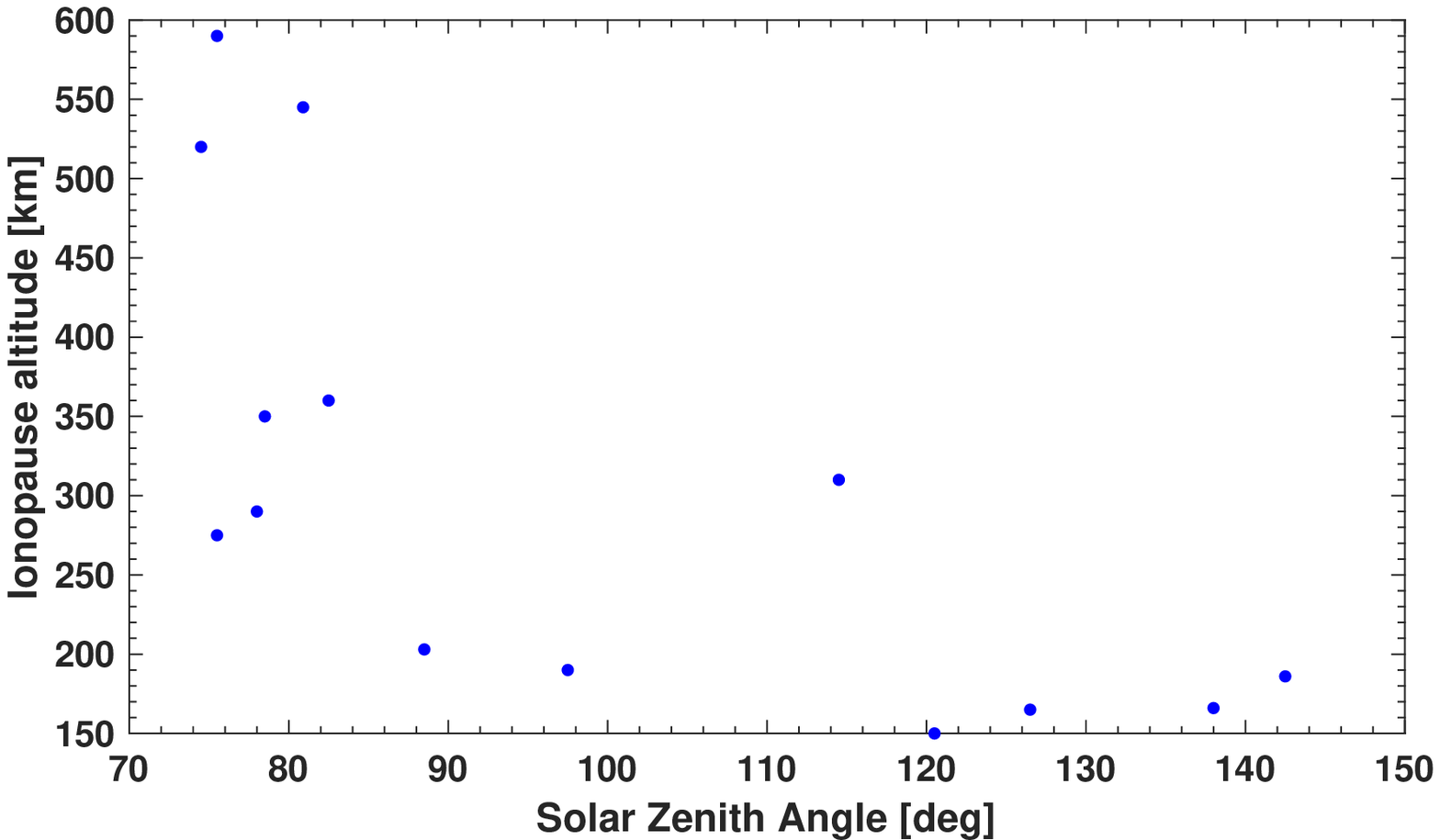}
	\caption{Electron density profiles of VEX 2014 egress occultations. The ionopause is placed between 150 km and 590 km, depending on the analyzed occultation. The ionopause altitude was chosen where the profile merged with the noise level \citep{Gerard2017}.}
	\label{fig7}
\end{figure}

Figure \ref{fig6} shows the electron density profile from the 1$^{st}$ February 2014 Venus Express egress occultation. As reported by \citet{Patzold2007}, Venus' ionosphere is characterized by two main electron density layers, identified as the secondary and main layers V1 and V2 respectively, which are clearly visible in Figure \ref{fig6}. In addition, the ionopause, which is the boundary between the solar wind flow and the planetary ionosphere, is placed at around 520 km altitude, where the electron density drops into the noise level. The analysis of the various ionospheric profiles obtained from VEX data of Figure \ref{fig7} suggests that the ionopause altitude varies between 150 km and 590 km due to the variability of the balance between the solar wind dynamic pressure and the ionospheric plasma pressure, as well as the solar zenith angle condition. Most of the profiles with lower SZA (DOY 020, 032, 036) experience higher ionopause altitudes than the night-time ones (DOY 060, 062, 064, 066). However, it is worth mentioning that, as presented by \citet{Gerard2017}, some single-frequency profiles could deviate with respect to the related dual-frequency ones (which are not available for this investigation due to the lack of S-band data for these occultations), due to "\textit{small-scale spacecraft bus vibrations or other vibrational disturbances at the ground station which produce an extra Doppler contribution along the line-of-sight and perturb the derived electron density profile}" \citep{Gerard2017}. For this reason, the ionopause altitudes could not be fully trusted. 

\begin{figure}
	\centering
	\includegraphics[scale=0.5]{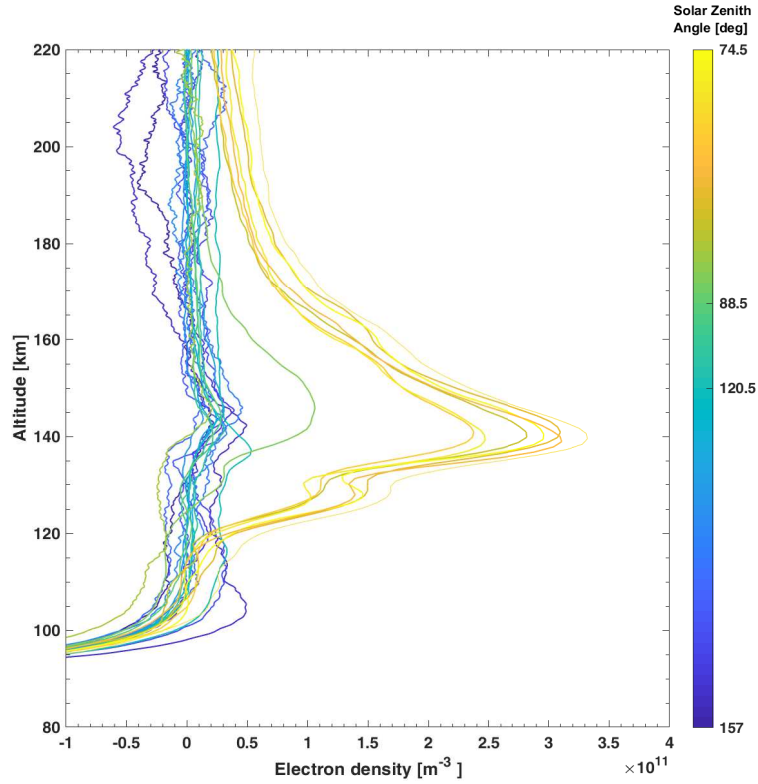}
	\caption{Electron density profiles of VEX occultations with day/night information.}
	\label{fig8}
\end{figure}

\begin{figure}
	\centering
	\includegraphics[scale=0.6]{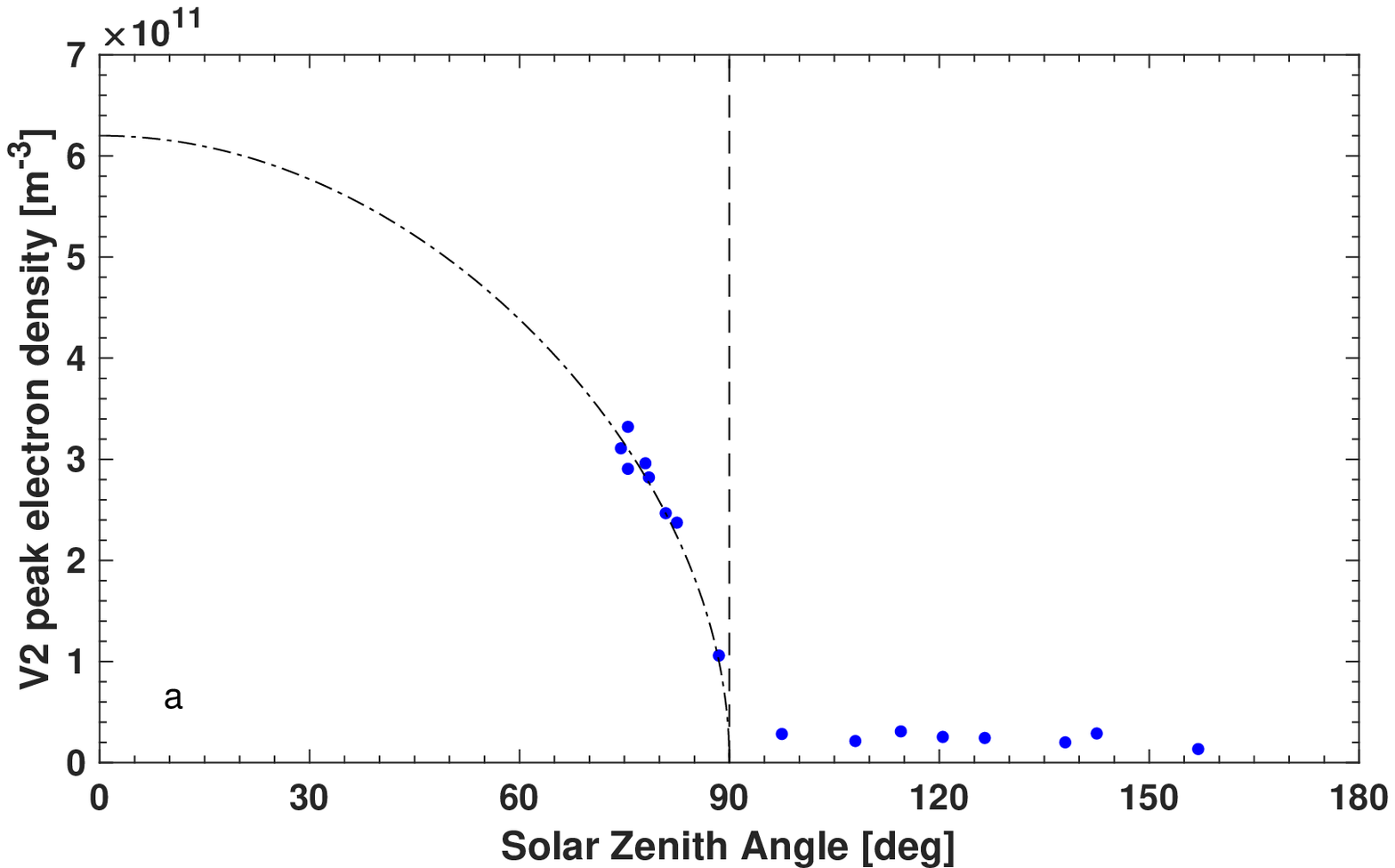}
	\centering
	\includegraphics[scale=0.6]{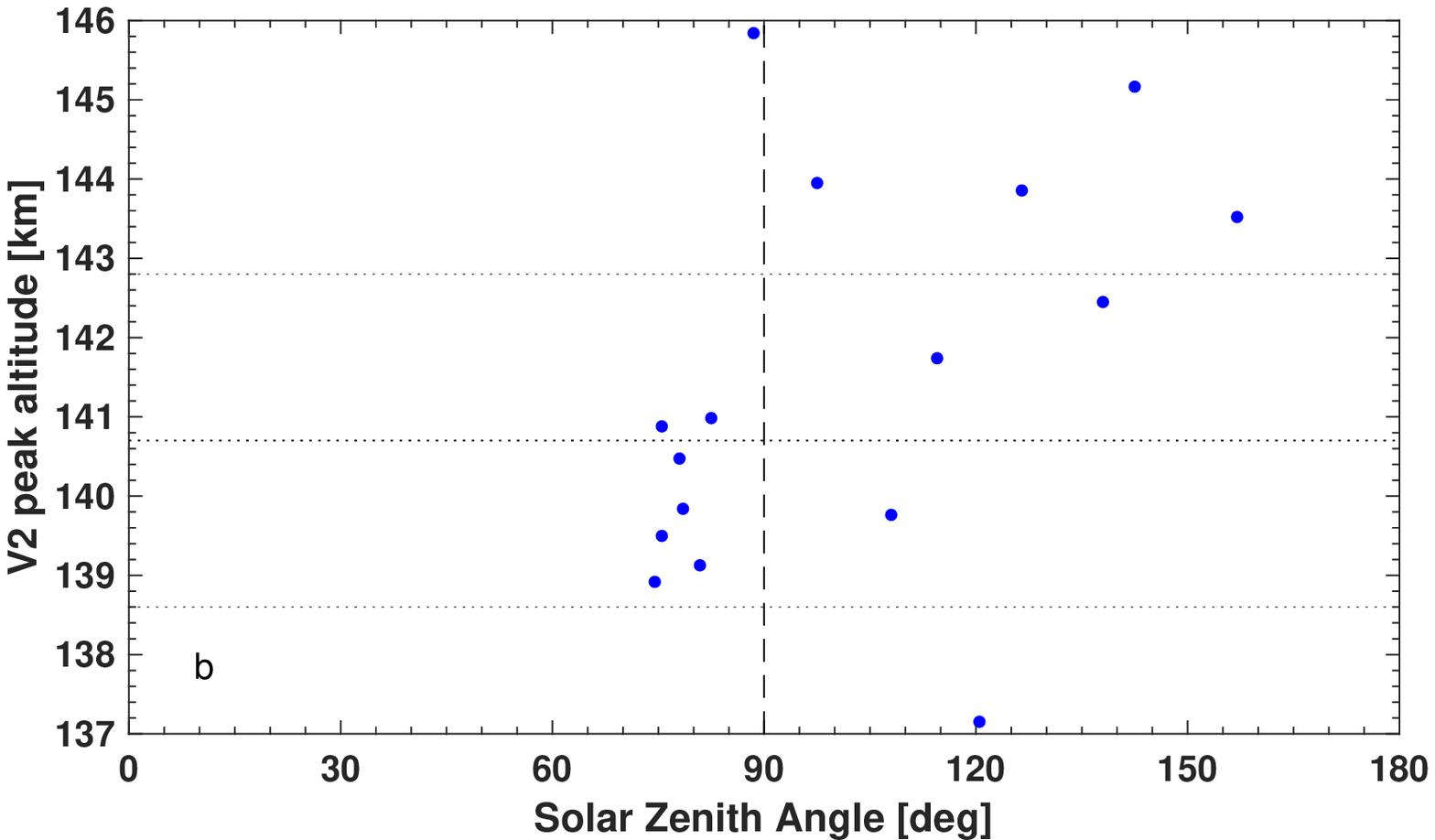}
	\caption{VeRa 2014 V2 data at solar maximum. The data matches the results reported by \citet{Gerard2017}. Panel (\textbf{a}): V2 peak density as a function of solar zenith angle. The dashed dotted curve is the ideal Chapman relation (see Eq. (7.3) of \citet{Gerard2017}). Panel (\textbf{b}): V2 peak altitudes. The peak altitude are constant for SZA $<=$ 80°, and they remain within the average $\pm$ 3 $\sigma$ found by \citet{Gerard2017}, which we report here as dotted lines (140.7 km $\pm$ 2.1 km). A V2 peak altitude increase for SZA $>=$ 80°, a decrease for 90° $<=$ SZA $<=$ 100°, and uncorrelated altitudes for SZA $>=$ 100° (due to random night-time structure) are found in our investigation too, and they match the results of \citet{Gerard2017}. }
	\label{v22}
\end{figure}

The ionosphere is ionized by direct solar radiation which charges up the ionosphere's particles, as a consequence a direct relation between electron density and solar zenith angle (day/night condition) is predictable and has been first presented by \citet{Ivanov-Kholodny1979}, \citet{Bauer1985} and more recently by \citet{Gerard2017} and \citet{Hensley2020}. Figure \ref{fig8} depicts the electron densities profiles for the analyzed 25 VEX occultations with additional information on the solar zenith angle and the results are in line with previous studies \citep{Patzold2007,Gerard2017}. Our profiles confirm the relation between the ionosphere and the day/night condition, highlighting the day-time occultations with a clearly defined electron density secondary and main layers V1 and V2, while the night-time occultations showing only the main layer V2, which is characterized by a lower electron density value. Furthermore, the night-time profiles of Figure \ref{fig8}, which have comparable electron density values, are characterized both by occultations occurred near the equator and in the polar regions. This suggests that Venus' ionosphere electron density is not influenced by the latitude and the dominant factor is the day/night condition.

To conclude, Figure \ref{v22} reports the typical Chapman plot of V2 peak altitudes and peak electron densities of Figure \ref{fig8} as a function of the solar zenith angle. Regarding the V2 peak electron densities, for Venus the \textit{Chapman} theory predicts in an excellent way the variation of the peak density and altitude with SZA, see Figure \ref{v22}. In general, our results are in agreement with the more comprehensive study carried out in \citet{Gerard2017}.

\subsection{Neutral atmosphere}
Radio signals are sensitive to Venus' neutral atmosphere approximately below 100 km altitude. Within this investigation the results are limited to an altitude of 38 km due to the critical refraction of Venus' atmosphere: the strong bending, refraction, absorption and defocusing caused by the atmosphere affect the radio signal, which becomes extremely weak and detection is no longer possible \citep{Eshleman1973}. As a result, at these altitudes radio occultation data are dominated by noise and the software is not able to retrieve the correct value of the bending angle. The lower boundary of our profiles is determined by the last frequency residual which yields a reliable bending angle parameter. The neutral number density \textit{n(h)} is evaluated from the neutral atmosphere refractivity $\nu_n$ through:

\begin{equation}
	n(h) = {\nu_n(h) \over k}\mathrm{,}
\end{equation}
where $k$ is the refractive volume of Venus' atmosphere \citep{Withers2010,Tellmann2009}.

The mass density is evaluated by assuming a known mean molecular mass $m_{mm}$ of the atmosphere \citep{nasa}:

\begin{equation}
	\rho(h) = n(h) \cdot m_{mm}\mathrm{.}
\end{equation}

\begin{figure}
	\centering
	\includegraphics[scale=0.5]{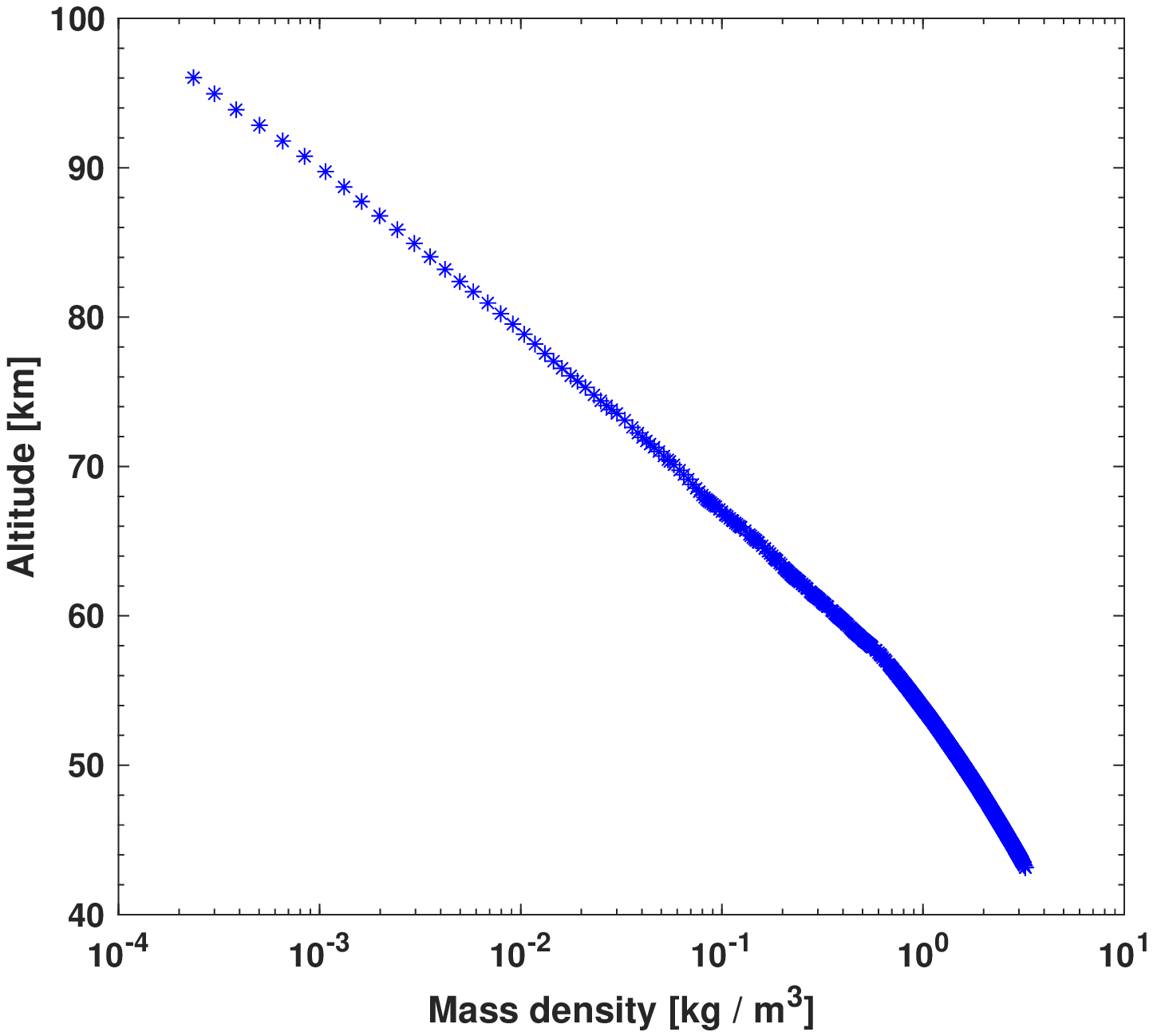}
	\includegraphics[scale=0.5]{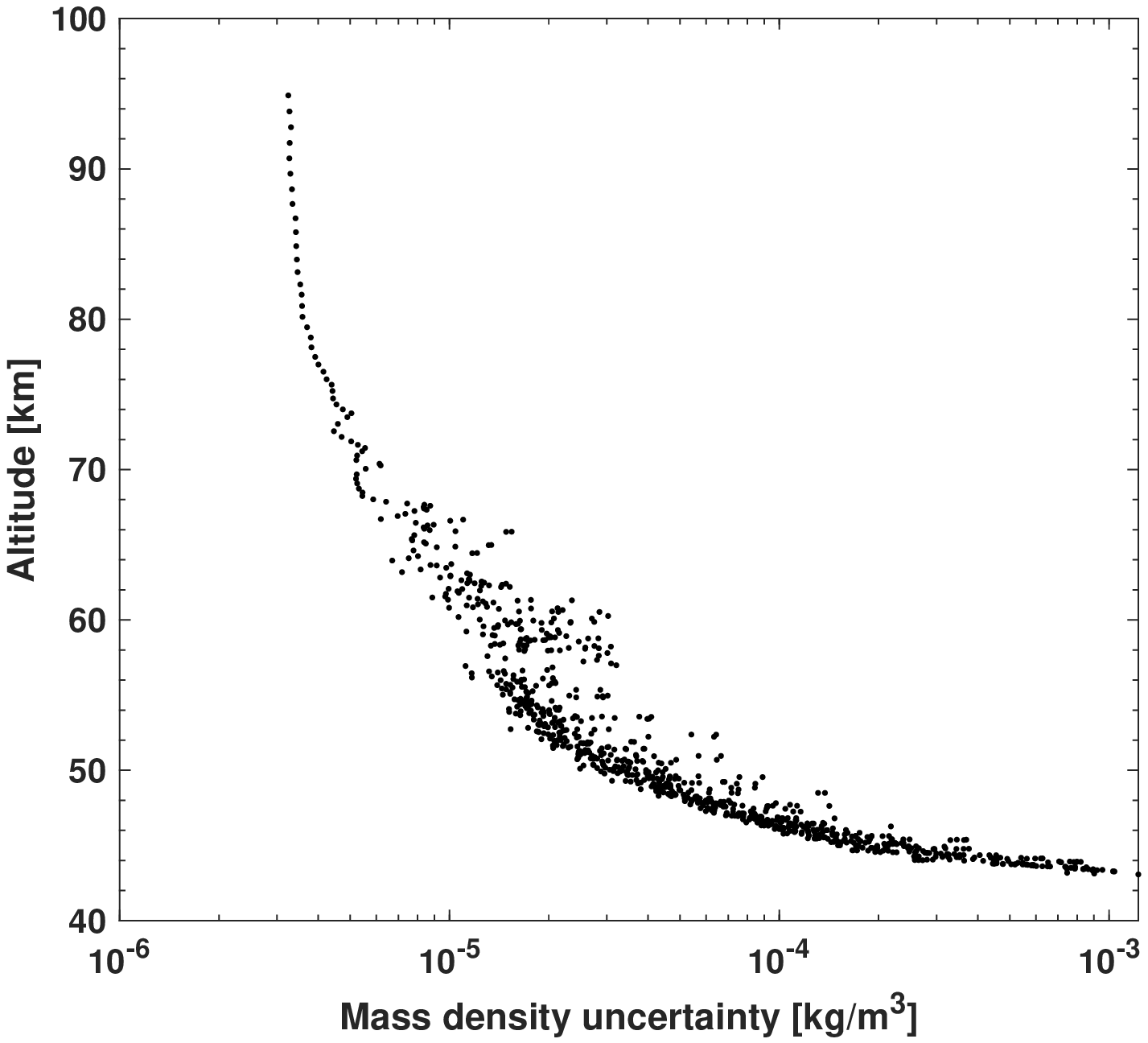}
	\centering
	\caption{Mass density profile of Venus' neutral atmosphere from VEX DOY 032 2014 egress occultation and related uncertainties. The uncertainties are retrieved from a Monte Carlo method, taking into account the noise present in the baseline of the frequency residuals, as well as a variable thermal noise function of altitude, see Appendix.}
	\label{fig10}
\end{figure}

Figure \ref{fig10} shows the mass density profile from the VEX DOY 032 2014 egress occultation and related uncertainties.

The temperature is obtained assuming hydrostatic equilibrium, through the method shown by \citet{Eshleman1973} and \citet{Tellmann2009}, by imposing a boundary condition $T_{up}$ at the upper boundary of the detectable atmosphere h$_{up}$ (100 km altitude):

\begin{equation}
	\label{eq4}
	T(h) = { \mu_{up} \over \mu(h)} \cdot T_{up} + { m_{mm} \over k \cdot n(h) } \int_h^{h_{up}} n(h') \cdot g(h') \mathrm{d}h' \mathrm{,}
\end{equation}
where $g(h)$ is the altitude-dependent acceleration of gravity.

Different boundary conditions do not influence the retrieved profiles below a certain altitude. For Venus the profiles converge below 85 km altitude regardless of the adopted boundary condition \citep{Patzold2007,Tellmann2009,Tellmann2012}.

\begin{figure}
	\centering
	\includegraphics[scale=0.5]{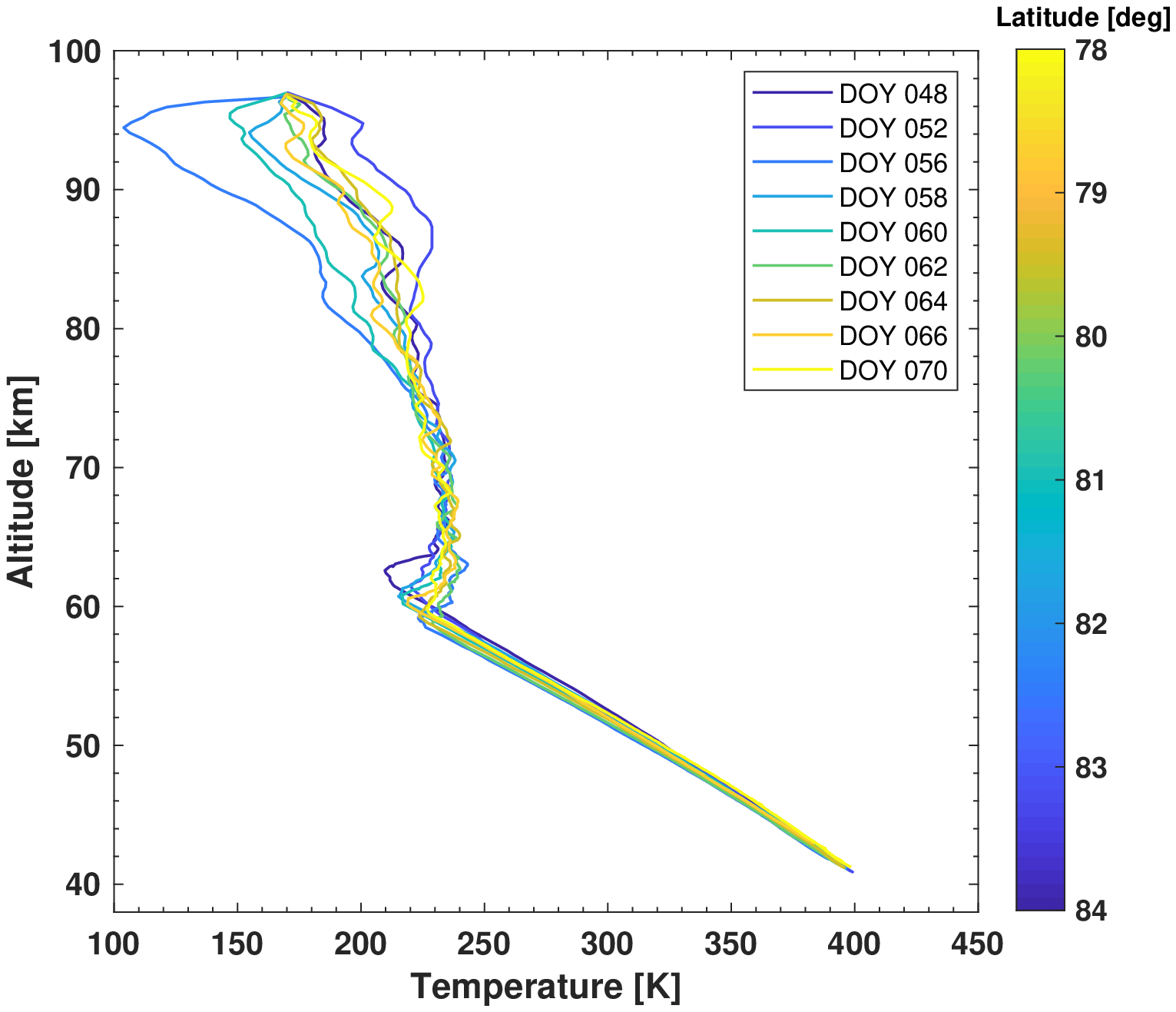}
    \includegraphics[scale=0.5]{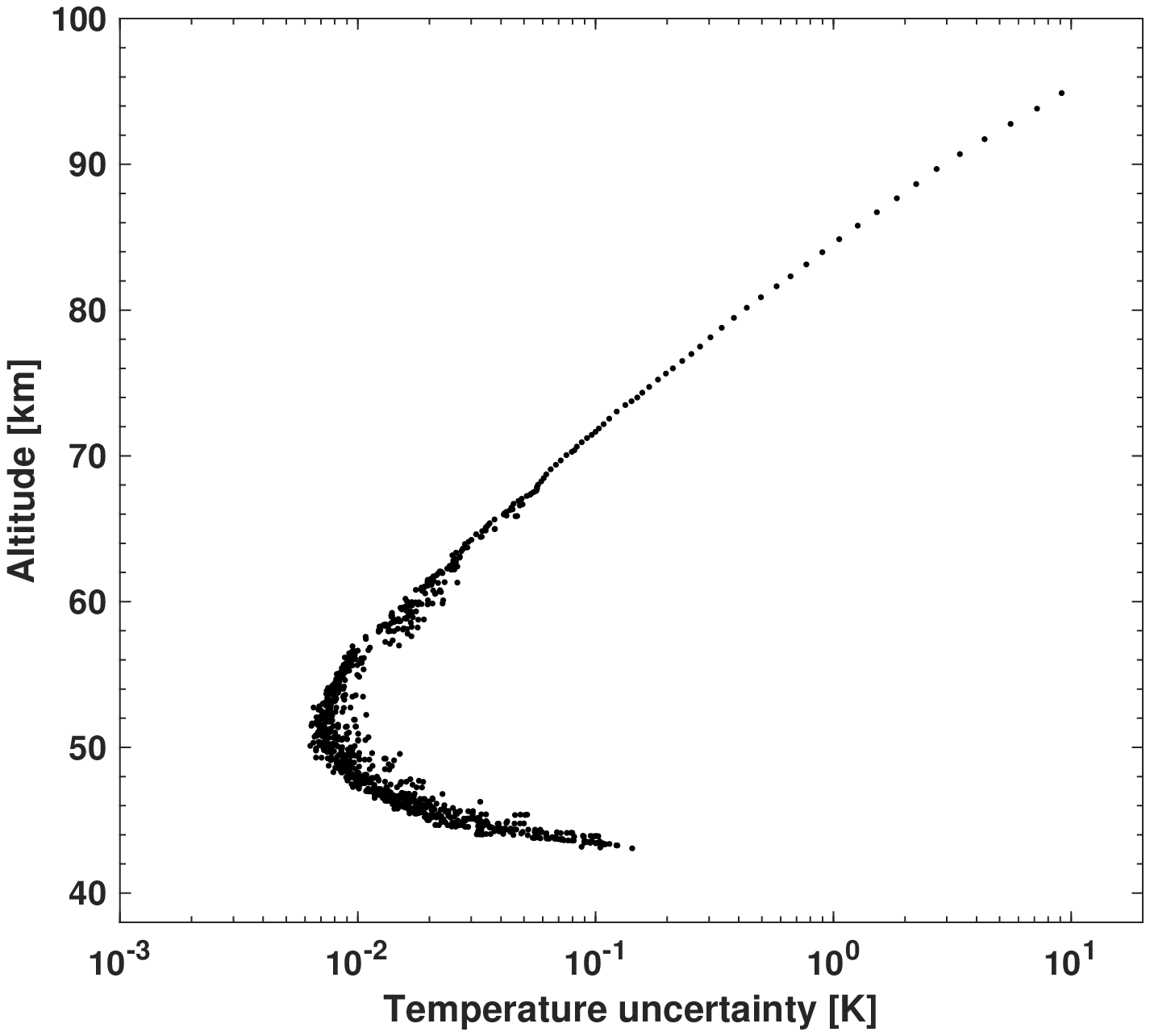}
	\centering
	\caption{Temperature profiles from VEX 2014 ingress occultations in the northern hemisphere and related uncertainties. The uncertainties are retrieved from a Monte Carlo method, taking into account the noise present in the baseline of the frequency residuals, as well as a variable thermal noise function of altitude and an assumed 20 K uncertainty in the boundary temperature, see Appendix.}
	\label{fig11}
\end{figure}

\begin{figure}
	\centering
	\includegraphics[scale=0.6]{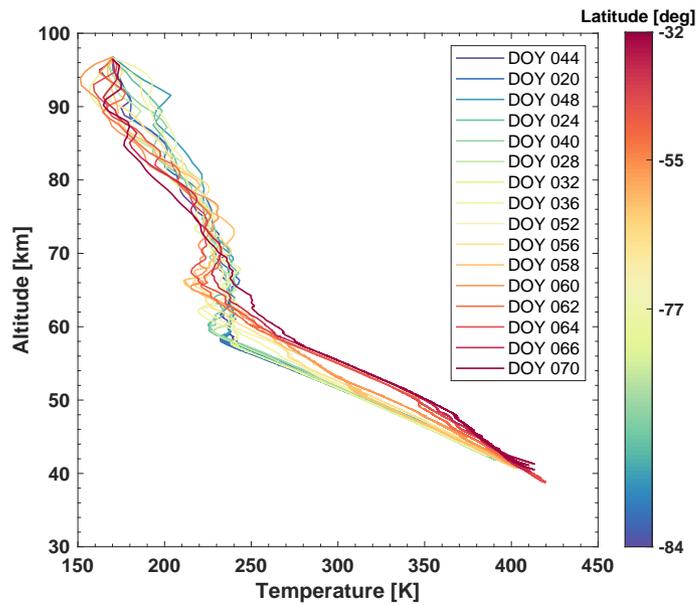}
	\centering
	\caption{Temperature profiles from VEX 2014 egress occultations in the southern hemisphere. The altitude is evaluated with respect to the mean Venus radius of 6051.8 km.}
	\label{fig12}
\end{figure}

In Figure \ref{fig11} and Figure \ref{fig12} we show the temperature profiles and related uncertainties obtained for the ingress and egress occultations, respectively. Our results are in agreement with previous studies by \citet{Patzold2007,Tellmann2009,Tellmann2012,Limaye2017}, showing that Venus experiences the same general temperature trend through decades. The maximum observed temperature is 420 K $\pm$ 0.1 K at 40 km altitude. We observe the same mean lapse rate of $\approx$ 10 K/km below the tropopause (around 60 km altitude) as presented in Figure 1 of \citet{Patzold2007,Tellmann2009}. Furthermore, the temperature profiles of Figure \ref{fig12} show a latitude-dependence, with over 40 K higher temperatures near the equator than at the poles below 60 km altitude \citep{Tellmann2009,Tellmann2012}. It is interesting also to note that the temperature difference between different latitudes tends to decrease and level out below 40 km altitudes.

\begin{figure}
	\centering
	\includegraphics[scale=0.4]{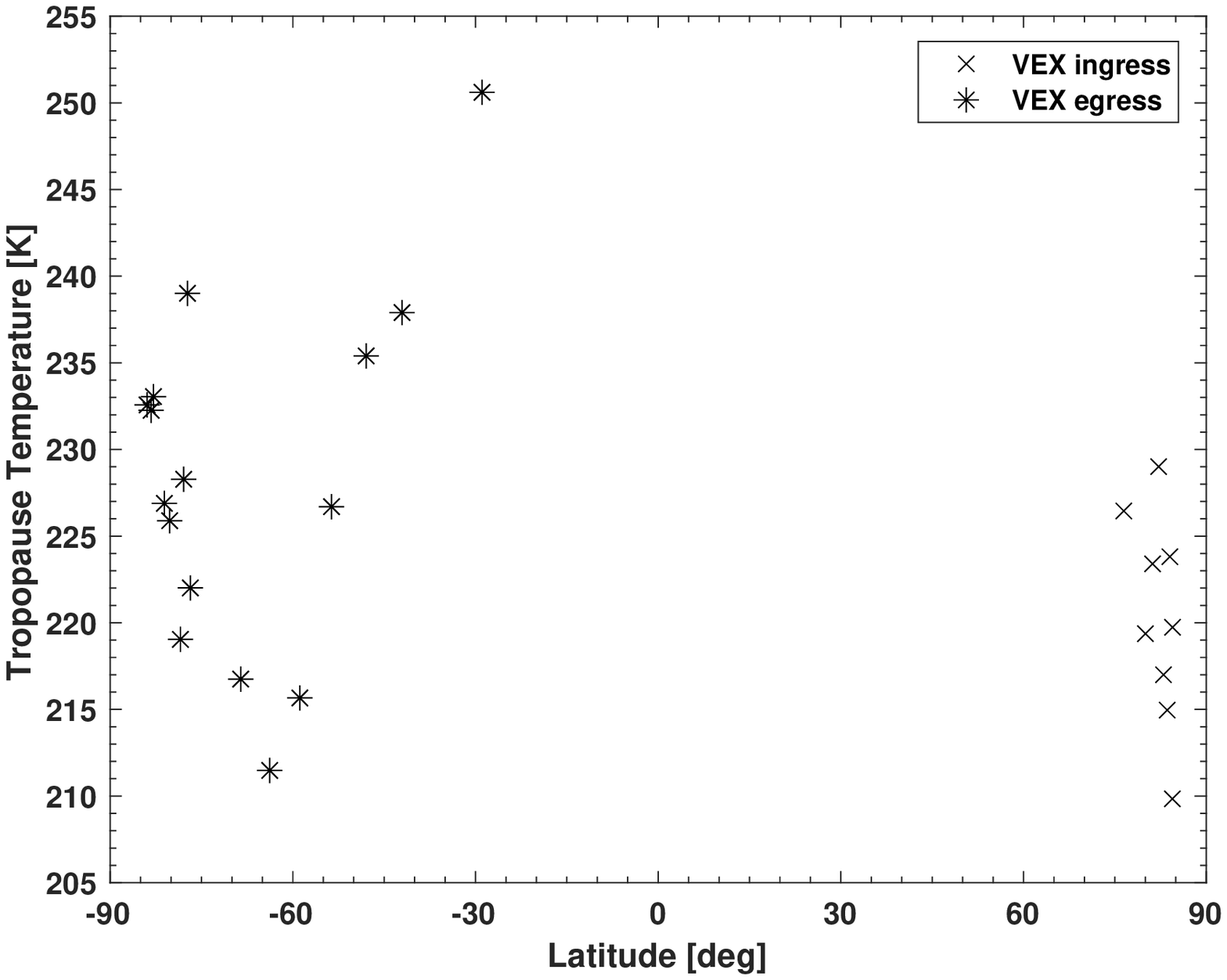}
	\includegraphics[scale=0.4]{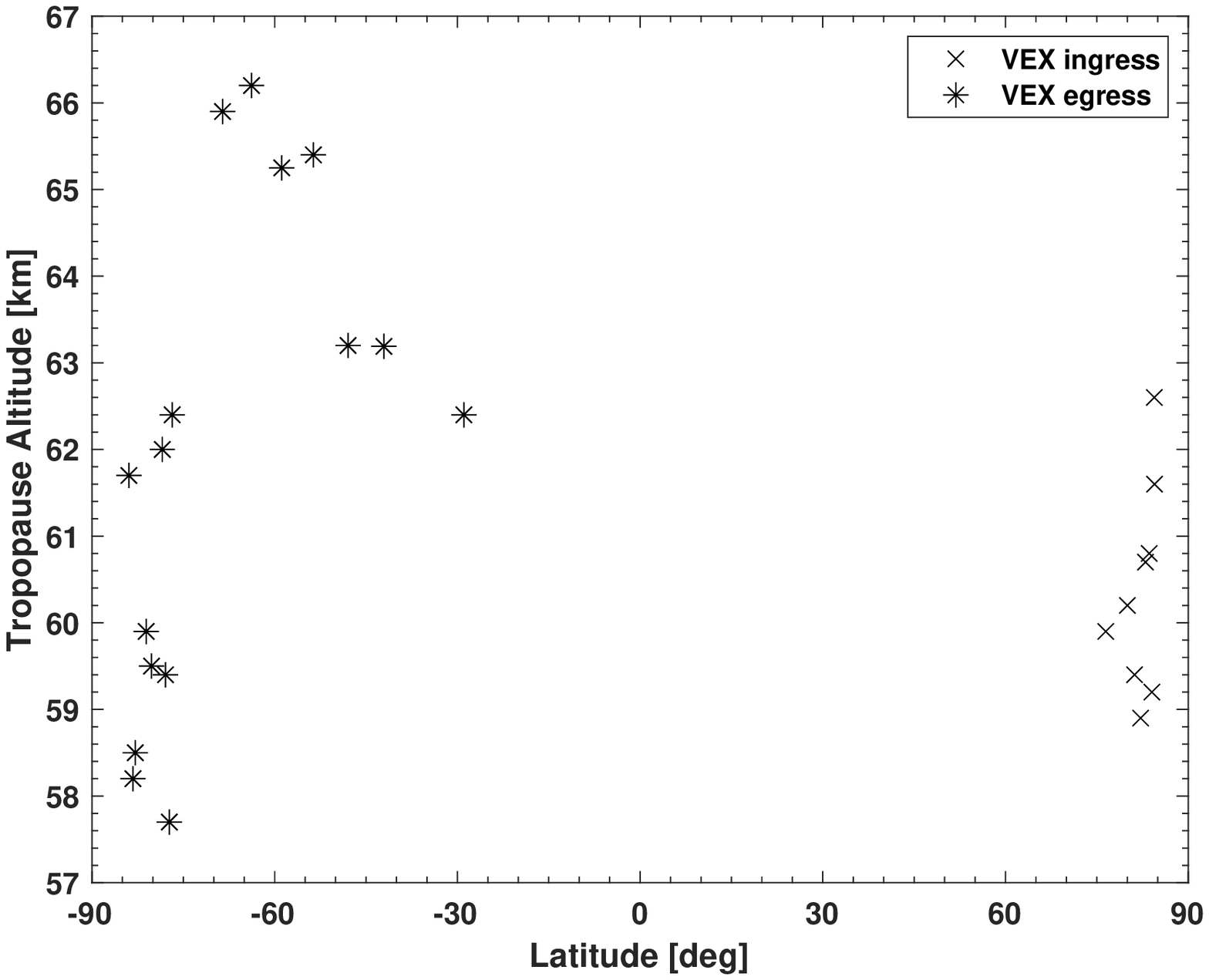}
	\caption{Venus' tropopause temperature and altitude vs latitude from VEX 2014 data.}
	\label{fig13}
\end{figure}

In addition, the tropopause, which is the temperature inversion layer which separates the mesosphere and the troposphere at around 60 km altitude, is quite clearly visible in the profiles. Since most of the temperature profiles retrieved in this work have a clear defined and easily identifiable tropopause, the tropopause altitudes and temperatures are sampled by simply looking at the different curves (a detailed study would consider also the analysis on the static stability of the atmosphere, which were not required within this work). The tropopause is found at altitudes between 57-67 km with temperatures which vary between 210 and 251 K, see Figure \ref{fig13}. This analysis highlights also the "cold collar" region at latitudes between 60-70$^{\circ}$, which is characterized by higher tropopause altitude and lower tropopause temperature as shown by \citet{Tellmann2009}. In general, our results agree well with Figures 6-7 of \citet{Tellmann2009}. In particular, the maximum tropopause temperature is found at low latitudes, -32 deg, then it decreases toward the cold collar region, and it increases again closer to the southern hemisphere pole \citep{Tellmann2009, Kliore1980}. Our results at the southern pole is about 232.7 K and it matches \citet{Tellmann2009} with their presented value of 233.6 $\pm 4.1$ K. Regarding the northern pole, our result is about 220 K, slightly lower than \citet{Tellmann2009} but within their uncertainty. Also, our results confirm what already noted in \citet{Tellmann2009}, that is the VeRa tropopause temperatures at the poles are sligtly lower than the ones reported by PV-ORO and Venera-15 -16 \citep{Kliore1980,Yakovlev1991}. However, the data from these missions agree well in the cold collar region.
The tropopause altitudes for the VEX 2014 occultations show, in general, good agreement with the results in \citet{Tellmann2009,Kliore1980,Yakovlev1991}. It is interesting to note that at the cold collar region and at the northern pole the tropopause altitude is found at 66 and 61 km, respectively, a few km higher than \citet{Tellmann2009} who reported 63 km and $\approx$ 58 km, respectively. To conclude, our 2014 results confirm that Venus' tropopause strongly depends on the latitude \citep{Tellmann2009,Kliore1980,Yakovlev1991}.

The pressure is the last evaluated atmospheric parameter, which is computed assuming an ideal gas behavior of the neutral atmosphere:
\begin{equation}
	p(h) = \rho(h) T(h) {k_B \over m_{mm}}\mathrm{,}
\end{equation}
where $k_B$ is the Boltzmann's constant.

\begin{figure}
	\centering
	\includegraphics[scale=0.5]{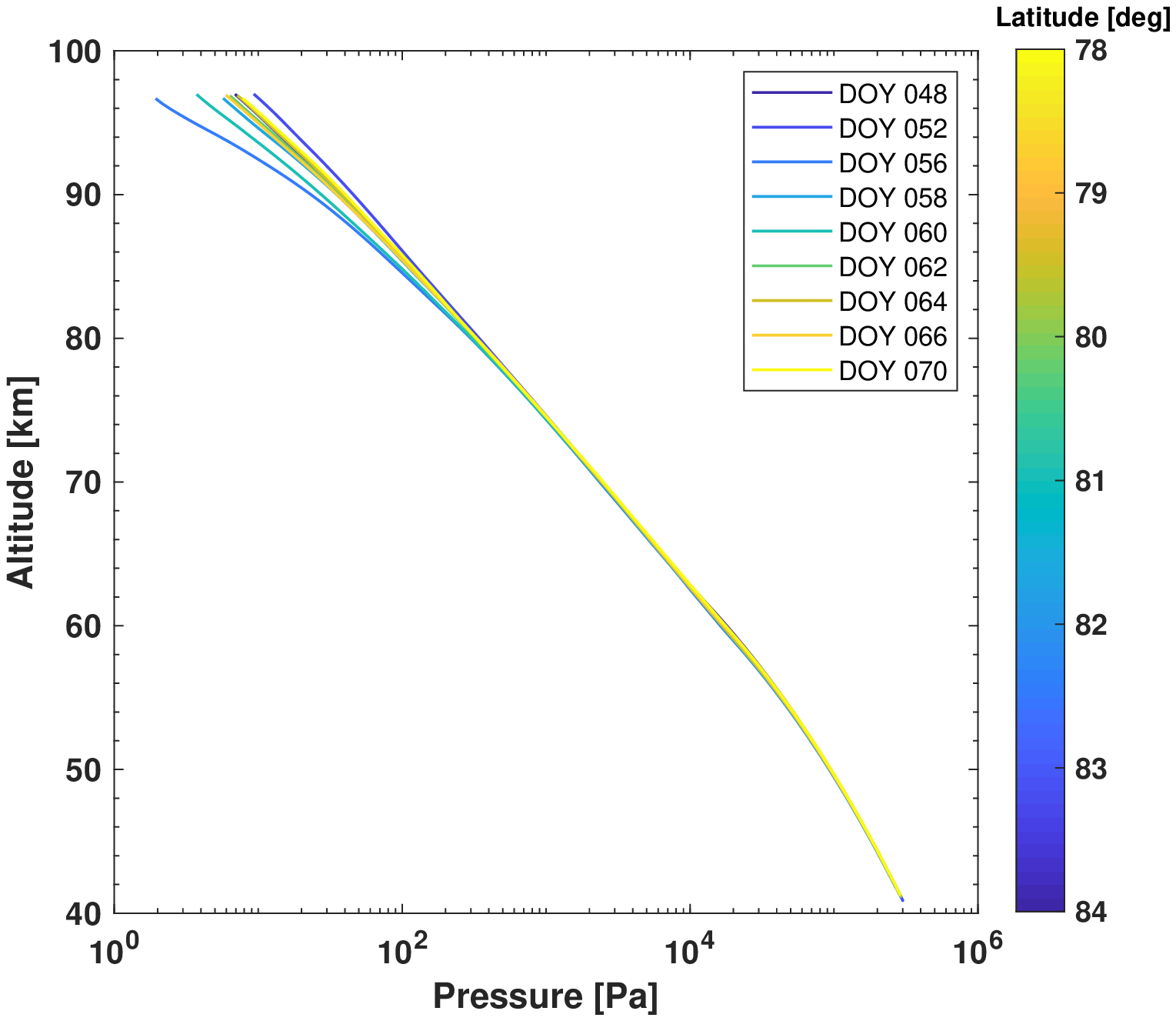}
	\includegraphics[scale=0.5]{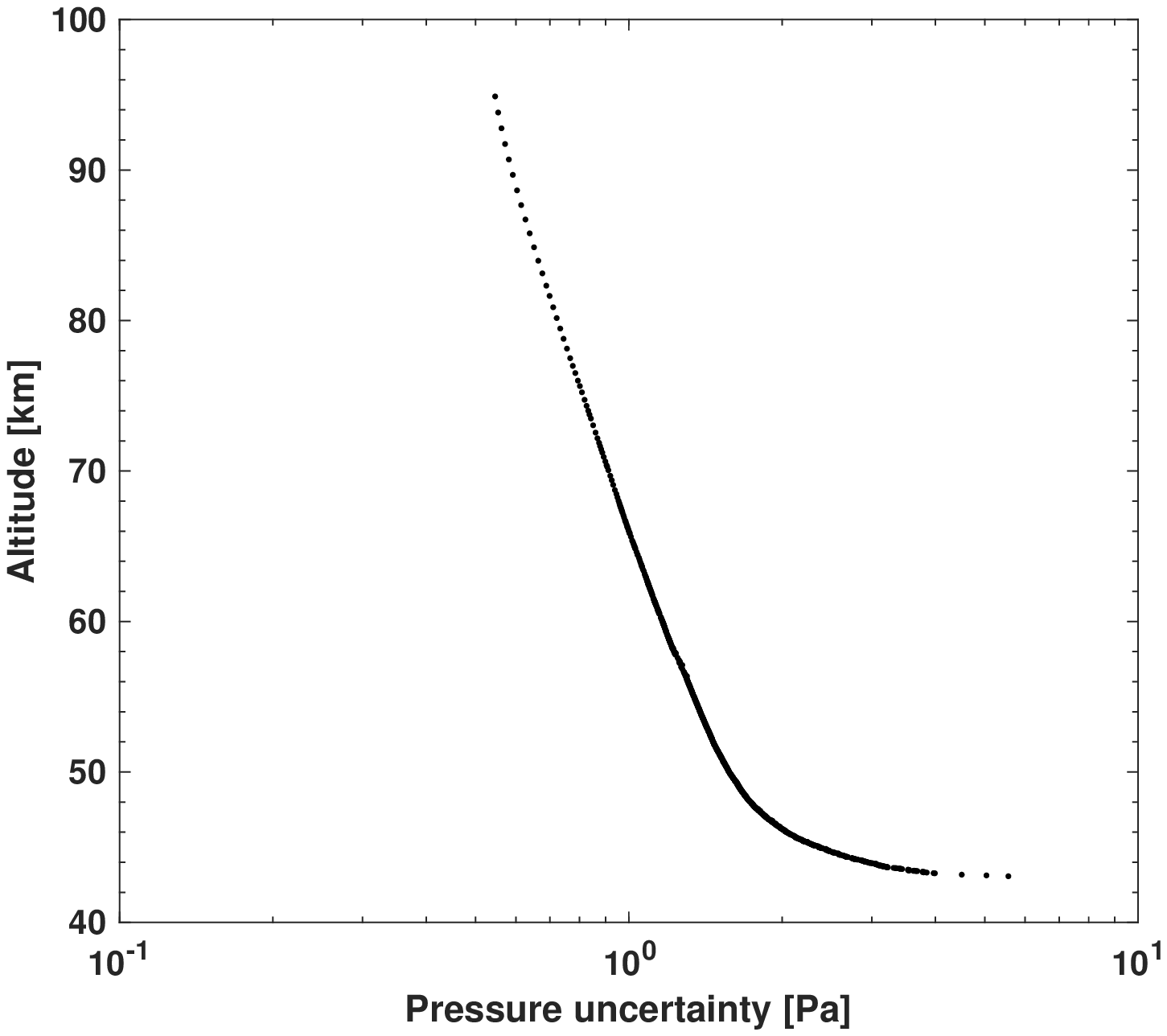}
	\centering
	\caption{Pressure profiles of the analyzed VEX 2014 ingress occultations and related uncertainties. The uncertainties are retrieved from a Monte Carlo method, taking into account the noise present in the baseline of the frequency residuals, as well as a variable thermal noise function of altitude, see Appendix.}
	\label{fig15}
\end{figure}

\begin{figure}
	\centering
	\includegraphics[scale=0.5]{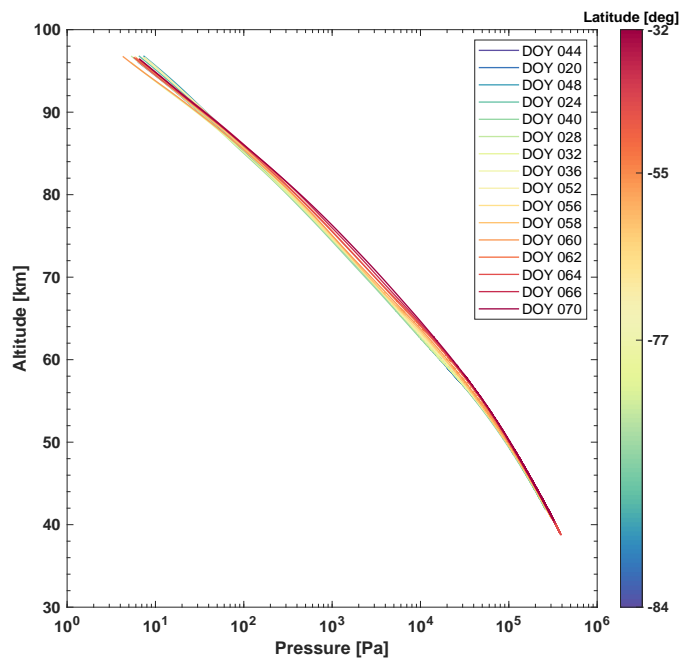}
	\caption{Pressure profiles of the analyzed VEX 2014 egress occultations.}
	\label{fig16}
\end{figure}

Figures \ref{fig15} and \ref{fig16} depict the pressure results, and related uncertainties, for the Venus Express ingress and egress cases, respectively.

As expected, since the profiles near the equator are characterized by higher temperatures they also show higher pressures than the polar regions. The maximum observed pressure is 4 bar $\pm$ 0.06 mbar.

\begin{figure}
	\centering
	\includegraphics[scale=0.5]{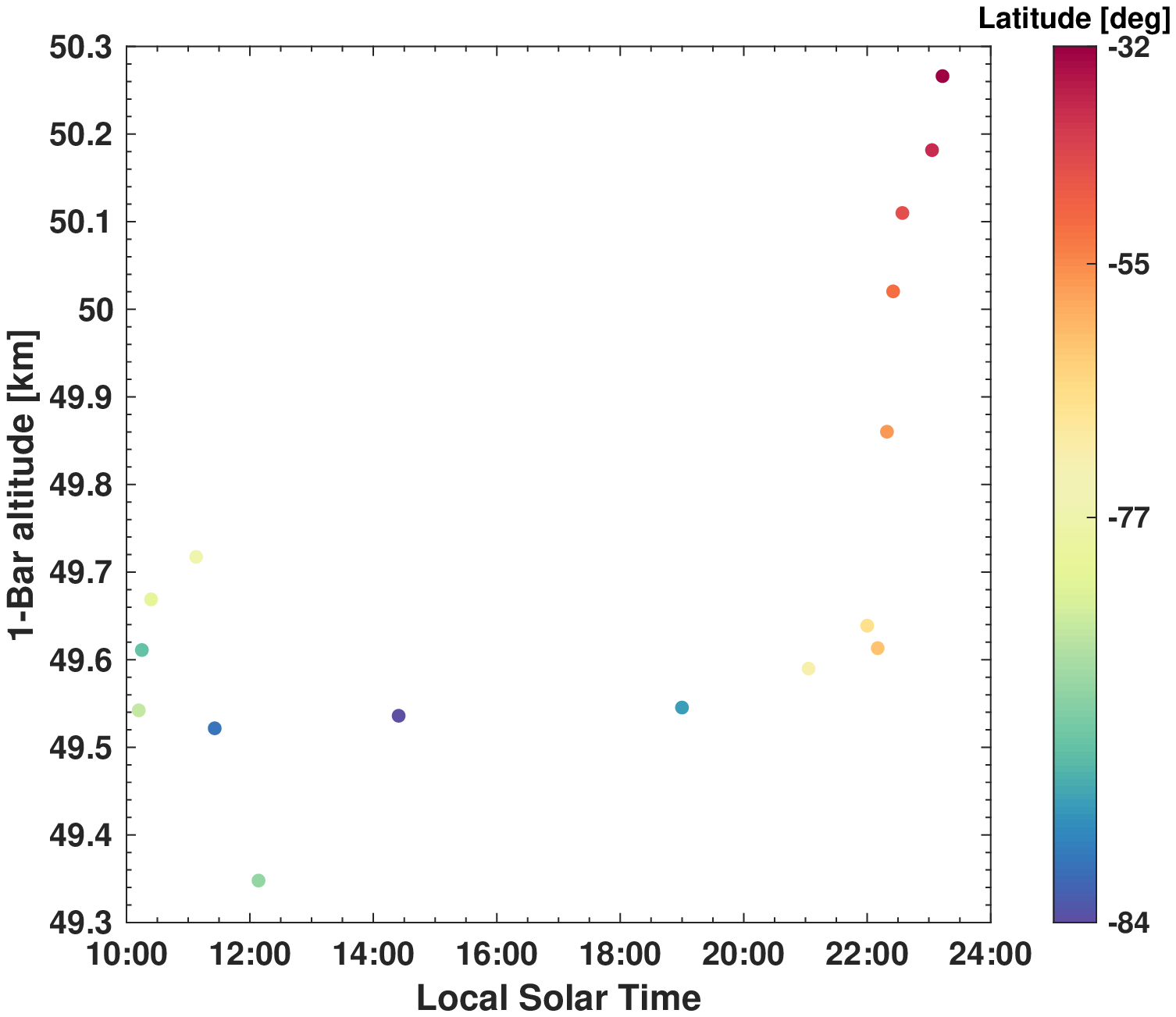}
	\includegraphics[scale=0.5]{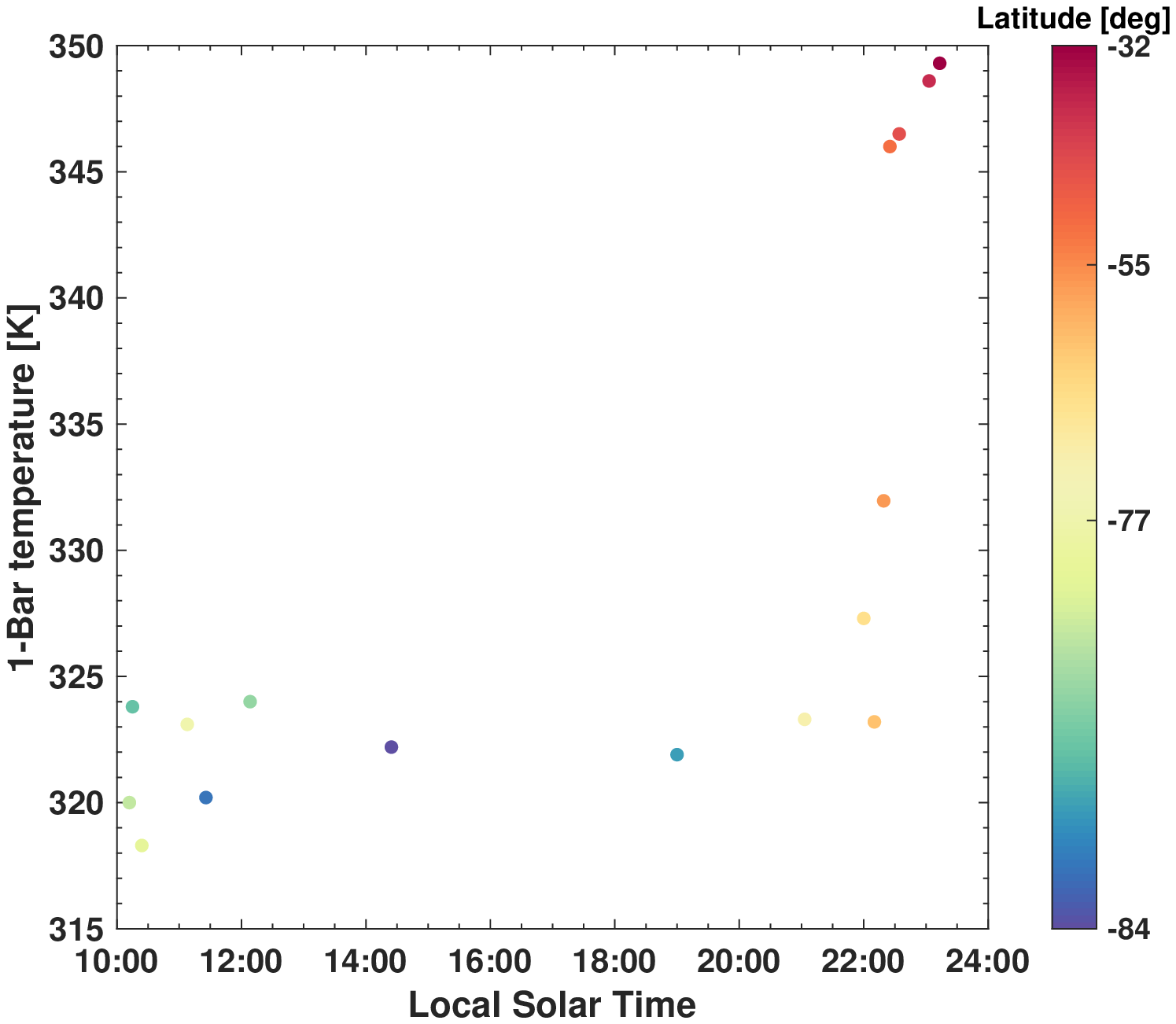}
	\centering
	\caption{1-bar altitudes and temperatures with day/night and latitude information from VEX 2014 data.}
	\label{fig17}
\end{figure}

To conclude, Figure \ref{fig17} shows the 1-bar altitudes and temperatures, as well as their dependences on latitude and day/night condition. VEX 2014 data show 1-bar temperatures between 320-350 K, and they perfectly match what found by \citet{Tellmann2009} in their Figure 11. Equatorial latitudes are characterized by higher temperatures and higher pressures than the polar regions, so a higher 1-bar altitude confirms these findings. However, the near-equator occultations probed Venus' atmosphere at local night-time. This result highlights that the near-equator regions in night-time are characterized by higher temperatures, pressures (and, therefore, 1 bar altitudes) than the day-time polar regions.

To summarize, the retrieved vertical profiles show that Venus' neutral atmosphere is influenced by the latitude but it is not affected by the day/night condition, in contrast with what observed in Venus' ionosphere, which is day/night-dependent and not latitude-dependent. These results are in agreement with previous studies presented in the literature \citep{Patzold2007,Tellmann2009,Tellmann2012,Limaye2017,Ivanov-Kholodny1979,Bauer1985,Hensley2020,Ando2020}. 

Our scientific interpretation of these results is related to two main factors, which take place within Venus' neutral atmosphere: Venus' \textit{high thermal inertia} and the presence of \textit{zonal winds} (which determine Venus' atmosphere super-rotation). At high altitudes, Venus' ionosphere is not affected by these effects and the day/night condition mainly determine a high or low electron density peak. On the other hand, at altitudes below 90 km, Venus' atmospheric rotational period is well known for being much faster than the one of the planet itself (the atmosphere takes 3-5 Earth’s days to complete a rotation around Venus, while one Venusian day is 243 Earth's days) and this super-rotation generates a strong and efficient heat transfer at mid-latitudes, all over the planet. The zonal winds are characterized by extremely high speeds, with mean velocities around 90 m/s and peaks as high as 140-150 m/s, especially at latitudes near the equator, which blow in the westward direction mainly. On the contrary, the meridional heat transfer (North-South and vice-versa) is less efficient than the zonal one \citep{Crisp1989,Hueso2012,Newman1984}. The second factor is the high thermal inertia of Venus, mostly due to the large amount of the atmosphere but also due to the greenhouse effect of its thick clouds, which traps the heat in the neutral atmosphere. We believe that the combination of these two factors can explain why the results show that the neutral atmosphere experiences mainly a latitude dependence rather than a day/night one. As a matter of fact, the mid-latitude zonal winds transfer the heat in a fast and efficient way, while at the same time the large amount of atmosphere, together with the greenhouse effect, trap the heat so that the equatorial regions at night-time do not suffer strong temperature differences with respect to the day-time ones. As a result, the occultation profiles acquired at night-time and at the equator are still characterized by higher 1-bar altitudes than those acquired near the poles.

\section{Conclusions}
This paper described the radio occultation methods and the required calibrations on raw data we used to generate vertical profiles of Venus' ionosphere and neutral atmosphere, starting from one-way, single frequency (X-band) signals from VEX-VeRa radio science instrument, recorded in open loop at the DSN stations. We described the data processing of our Abel inversion software, which includes all relativistic effects of the order of $1/c^2$. Moreover, we showed how the raw frequency residuals have been calibrated: in particular a first calibration was required to remove the effects of the local Earth's troposphere and ionosphere, while a second one, the so called baseline fit, is needed to compensate for any bias or linear trend in the frequency residuals caused by the spacecraft clock and estimated trajectory errors, in addition to plasma and thermal noise.

We analyzed 25 radio occultations carried out in 2014 with the goal to derive temperature and pressure vertical profiles, obtained from data received at a different ground station complex and with non-identical signal reception/processing methods than the principal VEX studies. The retrieved ionosphere vertical profiles are characterized by a ionopause found at altitudes between 150 and 590 km, which is a function of the balance between the solar wind dynamic pressure and the ionospheric plasma pressure but also of the solar zenith angle condition. In addition, at lower altitudes, our analysis shows Venus' tropopause between 57-67 km altitude, characterized by temperatures between 210 and 251 K, and we highlighted also the cold collar region at latitudes between 60-70$^{\circ}$, as already shown by \citet{Tellmann2009}. In general, as expected, our results are in agreement with the ones already available in the literature and they represent an additional validation of Venus' atmosphere results, derived from an independent analysis. In addition, the error analysis presents uncertainties which are comparable to the ones shown by \citet{Tellmann2012, Jenkins1994, Bocanegra2019} and \citet{Limaye2017}.

Furthermore, the analysis presented herein mainly focused on the influence of day/night condition and the latitude on Venus' atmosphere. The ionospheric profiles clearly showed a strong electron density dependence on the solar zenith angle rather than on the latitude, resulting in higher values of the electron density in the day-time regions than the night-time ones, regardless of the probed latitude. On the contrary, at lower altitudes the analysis of the neutral atmosphere highlighted temperature differences influenced mainly by the latitude of the occultation point, with the nigh-time equator regions characterized by higher temperatures than the day-time polar ones. However, these temperature differences tend to level out below 40 km altitude. 

Our scientific interpretation of these results suggests the high relevance of zonal winds and Venus' high thermal inertia, within Venus' neutral atmosphere. They determine a latitude dependence, rather than a day/night one, in the neutral atmosphere. The latter becomes dominant at high altitudes within Venus' ionosphere, where the two aforementioned effects are less important.

\section*{Appendix. Error Analysis}
\label{Appendix}
\renewcommand\thefigure{A\arabic{figure}} 
\renewcommand{\theHfigure}{A\arabic{figure}}
\setcounter{figure}{0}
In this appendix, we present the error analysis we performed in order to quantify the uncertainties of our retrieved profiles of Figures \ref{fig5}, \ref{fig10}, \ref{fig11}, \ref{fig15}. The uncertainties in the frequency time-series, $\sigma_{\Delta f}$, depend on several contributions, both random and systematic, as the adopted integration-time, the receiver' instrumentation noise, the transmitter frequency stability, the interplanetary plasma noise and the S/C ephemeris errors. These noises and errors are reflected in uncertainties in the derived profiles. To this end, Figures \ref{fig5}, \ref{fig10}, \ref{fig11}, \ref{fig15} consider three major sources of random error: the noise introduced by the spacecraft USO, $\sigma_{USO}$, the interplanetary plasma, $\sigma_{plasma}$, and the thermal noise introduced by the receiver at the DSN, $\sigma_{th}(a)$. The thermal noise is evaluated from Eq. 23 of \citet{Withers2010}, with $B=2$ Hz, $\tau_{thermal}=0.25$ s and a variable $C/N_0$, since the signal-to-noise ratio decreases with the altitude, the thermal noise increases at low altitudes. Let us emphasize that by $\sigma$ we refer to the Allan standard deviation of the noise source \citep{Allan1966}. Since we are dealing with single-frequency data, the noise due to the interplanetary plasma was estimated from the noise present in the baseline of the original data, after the baseline correction, by removing the expected thermal noise and USO noise:

\begin{equation}
	\label{sig1}
	\sigma_{plasma} = \sqrt{\sigma_{tot}^2-\sigma_{USO}^2-\sigma_{th}^2 } = 1.15 \cdot 10^{-12} \mathrm{,}
\end{equation}
where ${\sigma_{tot}} = 1.2 \cdot 10^{-12}$ is the Allan standard deviation of the noise present in the baseline of the VEX 032 2014 egress (used as a representative case) frequency residuals after the baseline correction, ${\sigma_{USO}} = 3 \cdot 10^{-13}$ \citep{Hausler2006}, ${\sigma_{th}} = 1.35 \cdot 10^{-13}$ is the Allan deviation of the thermal noise using the measured signal to noise ratio in the baseline.

Finally, our modeled $\sigma_{\Delta f}(a)$ is calculated following \citet{Bocanegra2019} and it includes:

\begin{equation}
	\label{sig2}
	\sigma_{\Delta f}(a) = \sqrt{\sigma_{USO}^2+\sigma_{plasma}^2+\sigma_{th}(a)^2 }\mathrm{,}
\end{equation}
where $\sigma_{th}(a) \cdot f$ ranges from $1.13$ mHz in the baseline to $226$ mHz at the last ray-path, and $f$ is the carrier frequency ( $\approx$ 8.4 GHz at X-band).


Our error analysis is based on a Monte Carlo estimation of the aforementioned errors in the atmospheric profiles, following \citet{Schinder2011,Schinder2012}. We performed 2000 runs (adding more profiles does not significantly change the results) of a representative NASA-DSN 2014 VEX occultation, by adding gaussian random noise time-series (whose Allan deviation is $\sigma_{\Delta f}(a)$) to the original frequency residuals $\Delta f(a)$, in order to obtain the standard deviations of the profiles in terms of relevant atmospheric parameters. In addition, the Monte Carlo includes an assumed 20 K uncertainty in the boundary temperature.

\begin{figure}
	\centering
	\includegraphics[scale=0.5]{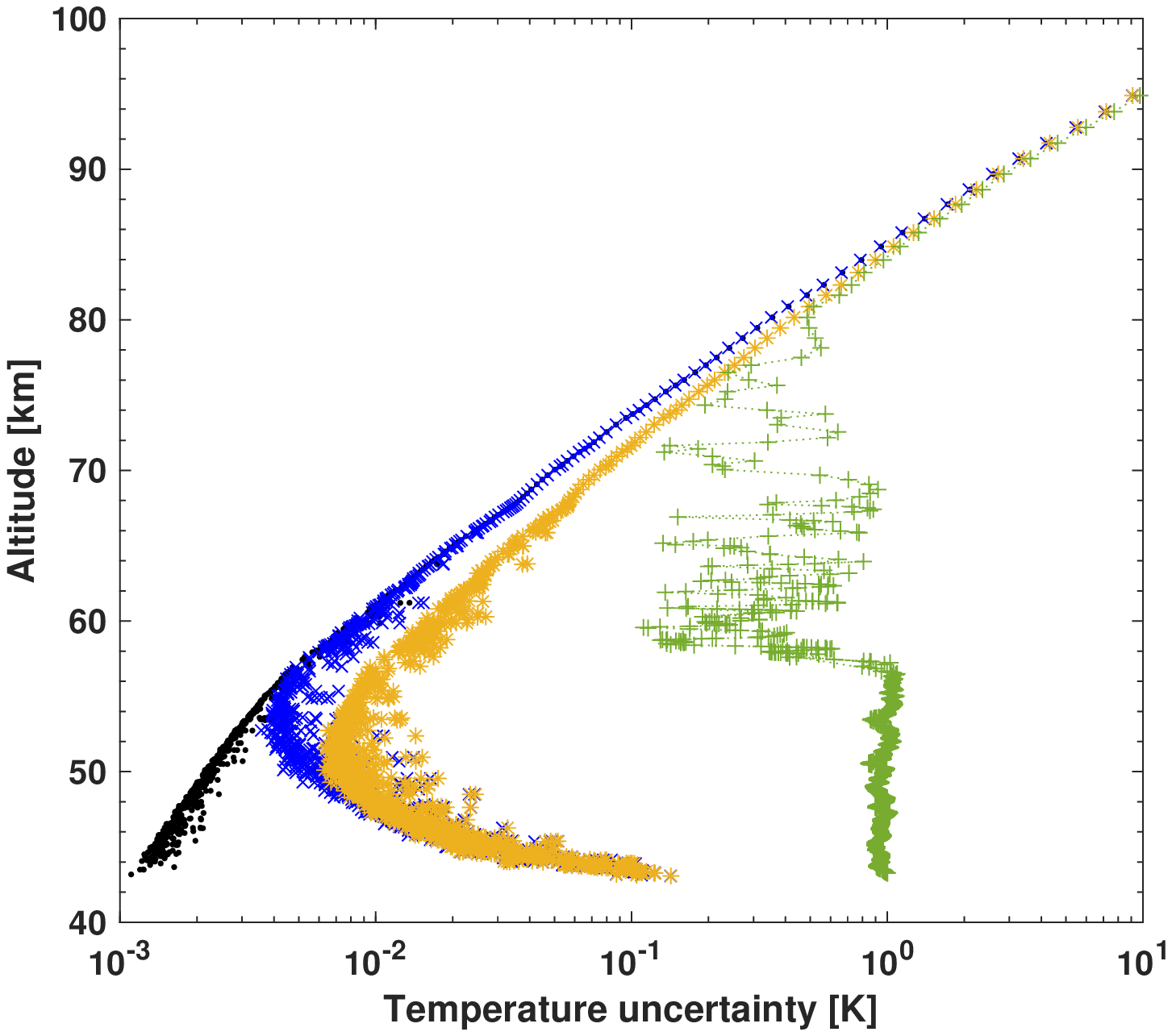}
	\includegraphics[scale=0.5]{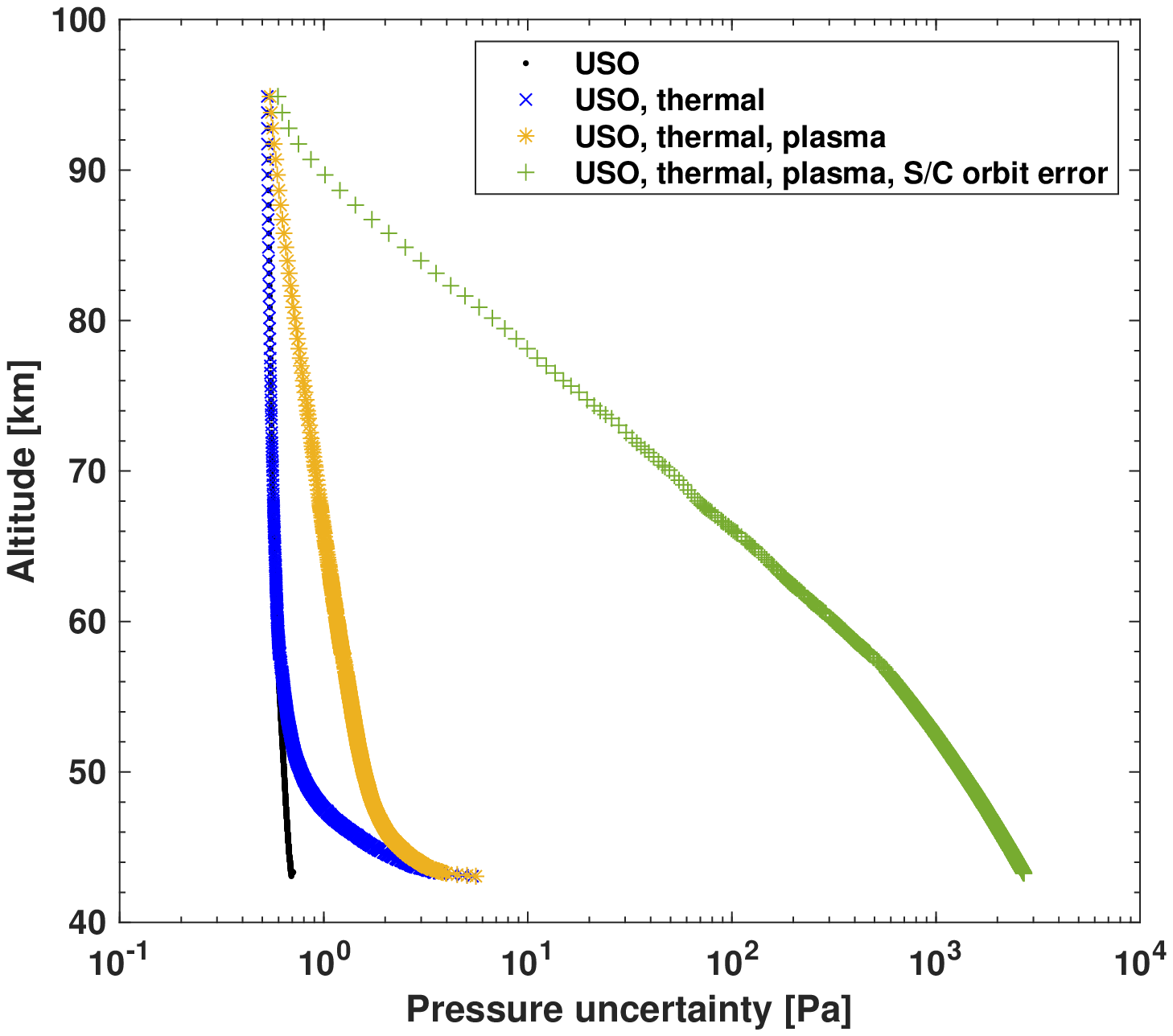}
	\centering
	\caption{Temperature and pressure uncertainties for the four error analysis. The data is from VEX DOY032 2014, acquired by the Deep Space Station DSS-45.}
	\label{fig18}
\end{figure}

Furthermore, it is worth mentioning that an error analysis is a function of the considered noise sources, so the resulting uncertainties will be different depending on the selected errors. We performed four error analyses, with the goal to show how different noise sources impact on the resulting uncertainties. The representative cases of these error analysis are summarized hereafter, and Figure \ref{fig18} shows the temperature and pressure uncertainties:

\begin{enumerate}
	\item USO noise, used as a comparison to \citet{Tellmann2012} to validate our error analysis;
	\item USO noise and thermal noise;
	\item USO noise, thermal noise, and plasma noise; this is the reference error analysis for the uncertainties presented in the manuscript and described above, and it is used as a comparison to \citet{Jenkins1994, Bocanegra2019} to validate our error analysis;
	\item USO noise, thermal noise, plasma noise and a systematic error in the spacecraft's position, to reproduce the spacecraft orbit inaccuracies. As reported by \citet{Tellmann2009}, the orbit inaccuracies of Venus Express are $\approx$ 100 m, which lead to errors in the altitude of the retrieved occultation profiles that are at the same level of 100 m, and so uncertainties in the temperatures and pressure profiles. So, we considered an uncertainty 1-sigma of 100 m in the spacecraft position along the projected path in the plane of the sky. This term has a strong effect especially in the last 20 km altitude, where the temperature gradient is about 10 K/km. A first-order evaluation shows that an error of 100 m in the spacecraft reconstructed orbit when probing that altitudes, results in an uncertainty of $10 K/km \cdot 0.1 km = 1K $, which is the value obtained in that region from our Monte Carlo error analysis, too.
\end{enumerate}
The values of the adopted USO, thermal and plasma Allan deviation contributions are the ones presented in Equations \ref{sig1}, \ref{sig2}. In addition, all the aforementioned error analysis assume 20 K uncertainty in the boundary temperature, too.

In order to validate our error analysis we adopted similar noise sources to the ones considered by previous studies as \citet{Tellmann2012}, \citet{Jenkins1994}, \citet{Bocanegra2019} and \citet{Limaye2017}. Our \textit{"USO"} temperature uncertainty curve in Figure \ref{fig18} is compared to the one in \citet{Tellmann2012}, while our \textit{"USO, thermal, plasma"} curve is compared to \citet{Bocanegra2019} and \citet{Jenkins1994}.\\ The agreement is very good in general: \citet{Tellmann2012} shows a temperature uncertainty of $10^{-3}$ K at 50 km altitude, while our error analysis shows $2 \cdot 10^{-3}$ K at the same altitude. 
Regarding \citet{Bocanegra2019}, their temperature and pressure uncertainties at about 48 km altitude are 0.015 K and 4 Pa, respectively, while our results show 0.01 K and 2 Pa at the same altitude. Moreover, the temperature uncertainty of \citet{Jenkins1994} is in the order of 0.1 K at 34 km altitude, and our result is 0.1 K at 43 km altitude. Regarding the pressures, \citet{Jenkins1994} shows 30 Pa at 44 km, while our analysis shows 5 Pa at the same altitude. The small discrepancies with respect to \citet{Jenkins1994} temperature and pressure uncertainties are linked to the use of different occultation data (VEX and Pioner Venus Orbiter) at different bands (VEX X-band and PVO S-band for \citet{Jenkins1994}), so the corresponding thermal noise and error sources are slightly different. In any case, both temperature and pressure curves shows the very same general trend, showing an increase of the uncertainties below 50 km altitude due to the thermal noise. To conclude, \citet{Limaye2017} in their Table 1 present VEX temperature uncertainties in the order of 0.1-1 K. This allows us to conclude that our error analysis, if considering similar noise sources, shows uncertainties in the same order of previous studies available in the literature.

\clearpage
\section*{Acknowledgements}
The Authors extend their most sincere thanks to Dr. \textit{Colin Wilson}, for his support in the first stages of the work. E.G., L.G.C., M.Z., and P.T. acknowledge sponsorship from the \textit{Alma Mater Studiorum - Università di Bologna}. This work was partially carried out at the \textit{Jet Propulsion Laboratory}, \textit{California Institute of Technology}, under a contract with the \textit{National Aeronautics and Space Administration}. Government sponsorship acknowledged. The Venus Express radio occultation data used in this paper was provided in 2019 by the Multi-mission "Planetary Radar and Radio Sciences Group" at Jet Propulsion Laboratory (JPL), California Institute of Technology. Finally, the  authors acknowledge the crucial contribution of the VeRa radio science team of the VEX mission, for planning the radio occultation experiments described in this paper, acquiring the data, publishing their results and then archiving the data in public repositories. Without their hard work and dedication, this paper would never have seen the light.

\bibliographystyle{model5-names}
\biboptions{authoryear}
\bibliography{references.bib}

\begin{thebibliography}{57}
\expandafter\ifx\csname natexlab\endcsname\relax\def\natexlab#1{#1}\fi
\providecommand{\url}[1]{\texttt{#1}}
\providecommand{\href}[2]{#2}
\providecommand{\path}[1]{#1}
\providecommand{\DOIprefix}{doi:}
\providecommand{\ArXivprefix}{arXiv:}
\providecommand{\URLprefix}{URL: }
\providecommand{\Pubmedprefix}{pmid:}
\providecommand{\doi}[1]{\href{http://dx.doi.org/#1}{\path{#1}}}
\providecommand{\Pubmed}[1]{\href{pmid:#1}{\path{#1}}}
\providecommand{\bibinfo}[2]{#2}
\ifx\xfnm\relax \def\xfnm[#1]{\unskip,\space#1}\fi
\bibitem[{Acton(1996)}]{Acton1996}
\bibinfo{author}{Acton, C.~H.} (\bibinfo{year}{1996}).
\newblock \bibinfo{title}{{Ancillary data services of NASA's navigation and
  Ancillary Information Facility}}.
\newblock {\it \bibinfo{journal}{Planetary and Space Science}\/},  {\it
  \bibinfo{volume}{44}\/}\bibinfo{issue}{(1 SPEC. ISS.)},
  \bibinfo{pages}{65--70}. \DOIprefix\doi{10.1016/0032-0633(95)00107-7}.
\bibitem[{Allan(1966)}]{Allan1966}
\bibinfo{author}{Allan, D.~W.} (\bibinfo{year}{1966}).
\newblock \bibinfo{title}{{Statistics of Atomic Frequency Standards}}.
\newblock {\it \bibinfo{journal}{Proceedings of the IEEE}\/},  {\it
  \bibinfo{volume}{54}\/}\bibinfo{issue}{(2)}, \bibinfo{pages}{221--230}.
  \DOIprefix\doi{10.1109/PROC.1966.4634}.
\bibitem[{Ando et~al.(2020)Ando, Imamura, Tellmann, P{\"{a}}tzold,
  H{\"{a}}usler, Sugimoto, Takagi, Sagawa, Limaye, Matsuda, Choudhary \&
  Antonita}]{Ando2020}
\bibinfo{author}{Ando, H.}, \bibinfo{author}{Imamura, T.},
  \bibinfo{author}{Tellmann, S.}, \bibinfo{author}{P{\"{a}}tzold, M.},
  \bibinfo{author}{H{\"{a}}usler, B.}, \bibinfo{author}{Sugimoto, N.},
  \bibinfo{author}{Takagi, M.}, \bibinfo{author}{Sagawa, H.},
  \bibinfo{author}{Limaye, S.}, \bibinfo{author}{Matsuda, Y.},
  \bibinfo{author}{Choudhary, R.~K.}, \& \bibinfo{author}{Antonita, M.}
  (\bibinfo{year}{2020}).
\newblock \bibinfo{title}{{Thermal structure of the Venusian atmosphere from
  the sub-cloud region to the mesosphere as observed by radio occultation}}.
\newblock {\it \bibinfo{journal}{Scientific Reports}\/},  {\it
  \bibinfo{volume}{10}\/}\bibinfo{issue}{(1)}, \bibinfo{pages}{1--7}.
  \DOIprefix\doi{10.1038/s41598-020-59278-8}.
\bibitem[{Barth(1967)}]{Barth1967}
\bibinfo{author}{Barth, C.~A.} (\bibinfo{year}{1967}).
\newblock \bibinfo{title}{{Venus: Ionosphere and atmosphere as measured by
  dual-frequency radio occultation of Mariner V}}.
\newblock {\it \bibinfo{journal}{Science}\/},  {\it
  \bibinfo{volume}{158}\/}\bibinfo{issue}{(3809)}, \bibinfo{pages}{1678--1683}.
  \DOIprefix\doi{10.1126/science.158.3809.1678}.
\bibitem[{Bauer et~al.(1985)Bauer, Brace, Taylor, Breus, Kliore, Knudsen, Nagy,
  Russell \& Savich}]{Bauer1985}
\bibinfo{author}{Bauer, S.~J.}, \bibinfo{author}{Brace, L.~H.},
  \bibinfo{author}{Taylor, H.~A.}, \bibinfo{author}{Breus, T.~K.},
  \bibinfo{author}{Kliore, A.~J.}, \bibinfo{author}{Knudsen, W.~C.},
  \bibinfo{author}{Nagy, A.~F.}, \bibinfo{author}{Russell, C.~T.}, \&
  \bibinfo{author}{Savich, N.~A.} (\bibinfo{year}{1985}).
\newblock \bibinfo{title}{{Chapter Vii the Venus Ionosphere}}.
\newblock {\it \bibinfo{journal}{Advances in Space Research}\/},  {\it
  \bibinfo{volume}{5}\/}\bibinfo{issue}{(11)}, \bibinfo{pages}{233--267}.
\bibitem[{Besse et~al.(2018)Besse, Vallat, Barthelemy, Coia, Costa, {De
  Marchi}, Fraga, Grotheer, Heather, Lim, Martinez, Arviset, Barbarisi,
  Docasal, Macfarlane, Rios, Saiz \& Vallejo}]{BESSE2018131}
\bibinfo{author}{Besse, S.}, \bibinfo{author}{Vallat, C.},
  \bibinfo{author}{Barthelemy, M.}, \bibinfo{author}{Coia, D.},
  \bibinfo{author}{Costa, M.}, \bibinfo{author}{{De Marchi}, G.},
  \bibinfo{author}{Fraga, D.}, \bibinfo{author}{Grotheer, E.},
  \bibinfo{author}{Heather, D.}, \bibinfo{author}{Lim, T.},
  \bibinfo{author}{Martinez, S.}, \bibinfo{author}{Arviset, C.},
  \bibinfo{author}{Barbarisi, I.}, \bibinfo{author}{Docasal, R.},
  \bibinfo{author}{Macfarlane, A.}, \bibinfo{author}{Rios, C.},
  \bibinfo{author}{Saiz, J.}, \& \bibinfo{author}{Vallejo, F.}
  (\bibinfo{year}{2018}).
\newblock \bibinfo{title}{Esa's planetary science archive: Preserve and present
  reliable scientific data sets}.
\newblock {\it \bibinfo{journal}{Planetary and Space Science}\/},  {\it
  \bibinfo{volume}{150}\/}, \bibinfo{pages}{131--140}. \URLprefix
  \url{https://www.sciencedirect.com/science/article/pii/S0032063316304688}.
  \DOIprefix\doi{https://doi.org/10.1016/j.pss.2017.07.013}.
\newblock \bibinfo{note}{Enabling Open and Interoperable Access to Planetary
  Science and Heliophysics Databases and Tools}.
\bibitem[{Bocanegra-Baham{\'{o}}n et~al.(2019)Bocanegra-Baham{\'{o}}n, {Molera
  Calv{\'{e}}s}, Gurvits, Cim{\`{o}}, Dirkx, Duev, Pogrebenko, Rosenblatt,
  Limaye, Cui, Li, Kondo, Sekido, Mikhailov, Kharinov, Ipatov, Wang, Zheng, Ma,
  Lovell \& McCallum}]{Bocanegra2019}
\bibinfo{author}{Bocanegra-Baham{\'{o}}n, T.~M.}, \bibinfo{author}{{Molera
  Calv{\'{e}}s}, G.}, \bibinfo{author}{Gurvits, L.~I.},
  \bibinfo{author}{Cim{\`{o}}, G.}, \bibinfo{author}{Dirkx, D.},
  \bibinfo{author}{Duev, D.~A.}, \bibinfo{author}{Pogrebenko, S.~V.},
  \bibinfo{author}{Rosenblatt, P.}, \bibinfo{author}{Limaye, S.},
  \bibinfo{author}{Cui, L.}, \bibinfo{author}{Li, P.}, \bibinfo{author}{Kondo,
  T.}, \bibinfo{author}{Sekido, M.}, \bibinfo{author}{Mikhailov, A.~G.},
  \bibinfo{author}{Kharinov, M.~A.}, \bibinfo{author}{Ipatov, A.~V.},
  \bibinfo{author}{Wang, W.}, \bibinfo{author}{Zheng, W.}, \bibinfo{author}{Ma,
  M.}, \bibinfo{author}{Lovell, J.~E.}, \& \bibinfo{author}{McCallum, J.~N.}
  (\bibinfo{year}{2019}).
\newblock \bibinfo{title}{{Venus Express radio occultation observed by PRIDE}}.
\newblock {\it \bibinfo{journal}{Astronomy and Astrophysics}\/},  {\it
  \bibinfo{volume}{624}\/}, \bibinfo{pages}{1--14}.
  \DOIprefix\doi{10.1051/0004-6361/201833160}.
\bibitem[{Born \& Wolf(1959)}]{Born1959}
\bibinfo{author}{Born, M.}, \& \bibinfo{author}{Wolf, E.}
  (\bibinfo{year}{1959}).
\newblock \bibinfo{title}{{Principles of Optics}}.
\bibitem[{{Bourgoin} et~al.(2021){Bourgoin}, {Zannoni}, {Gomez Casajus},
  {Tortora} \& {Teyssandier}}]{2021A&A...648A..46B}
\bibinfo{author}{{Bourgoin}, A.}, \bibinfo{author}{{Zannoni}, M.},
  \bibinfo{author}{{Gomez Casajus}, L.}, \bibinfo{author}{{Tortora}, P.}, \&
  \bibinfo{author}{{Teyssandier}, P.} (\bibinfo{year}{2021}).
\newblock \bibinfo{title}{{Relativistic modeling of atmospheric occultations
  with time transfer functions}}.
\newblock {\it \bibinfo{journal}{Astronomy and Astrophysics}\/},  {\it
  \bibinfo{volume}{648}\/}, \bibinfo{pages}{A46}.
  \DOIprefix\doi{10.1051/0004-6361/202040269}.
  \href{http://arxiv.org/abs/2012.15768}{\tt arXiv:2012.15768}.
\bibitem[{Crisp(1989)}]{Crisp1989}
\bibinfo{author}{Crisp, D.} (\bibinfo{year}{1989}).
\newblock \bibinfo{title}{{Radiative forcing of the Venus mesosphere. II.
  Thermal fluxes, cooling rates, and radiative equilibrium temperatures}}.
\newblock {\it \bibinfo{journal}{Icarus}\/},  {\it
  \bibinfo{volume}{77}\/}\bibinfo{issue}{(2)}, \bibinfo{pages}{391--413}.
  \DOIprefix\doi{10.1016/0019-1035(89)90096-1}.
\bibitem[{Dalba \& Withers(2019)}]{Dalba2019}
\bibinfo{author}{Dalba, P.~A.}, \& \bibinfo{author}{Withers, P.}
  (\bibinfo{year}{2019}).
\newblock \bibinfo{title}{{Cassini Radio Occultation Observations of Titan's
  Ionosphere: The Complete Set of Electron Density Profiles}}.
\newblock {\it \bibinfo{journal}{Journal of Geophysical Research: Space
  Physics}\/},  {\it \bibinfo{volume}{124}\/}\bibinfo{issue}{(1)},
  \bibinfo{pages}{643--660}. \DOIprefix\doi{10.1029/2018JA025693}.
\bibitem[{Eshleman(1973)}]{Eshleman1973}
\bibinfo{author}{Eshleman, V.~R.} (\bibinfo{year}{1973}).
\newblock \bibinfo{title}{{The radio occultation method for the study of
  planetary atmospheres}}.
\newblock {\it \bibinfo{journal}{Planetary and Space Science}\/},  {\it
  \bibinfo{volume}{21}\/}\bibinfo{issue}{(9)}, \bibinfo{pages}{1521--1531}.
  \DOIprefix\doi{10.1016/0032-0633(73)90059-7}.
\bibitem[{Fjeldbo \& Eshleman(1969)}]{Fjeldbo1969}
\bibinfo{author}{Fjeldbo, G.}, \& \bibinfo{author}{Eshleman, V.~R.}
  (\bibinfo{year}{1969}).
\newblock \bibinfo{title}{{Atmosphere of Venus as Studied with the Mariner 5
  Dual Radio‐Frequency Occultation Experiment}}.
\newblock {\it \bibinfo{journal}{Radio Science}\/},  {\it
  \bibinfo{volume}{4}\/}\bibinfo{issue}{(10)}, \bibinfo{pages}{879--897}.
  \DOIprefix\doi{10.1029/RS004i010p00879}.
\bibitem[{Fjeldbo et~al.(1971)Fjeldbo, Kliore \& Eshleman}]{Academy1971}
\bibinfo{author}{Fjeldbo, G.}, \bibinfo{author}{Kliore, A.~J.}, \&
  \bibinfo{author}{Eshleman, V.~R.} (\bibinfo{year}{1971}).
\newblock \bibinfo{title}{{The Neutral Atmosphere of Venus as Studied with the
  Mariner V Radio Occultation Experiments}}.
\newblock {\it \bibinfo{journal}{The Astronomical journal}\/},  {\it
  \bibinfo{volume}{225}\/}\bibinfo{issue}{(2)}, \bibinfo{pages}{1--21}.
  \DOIprefix\doi{10.1086/111096}.
\bibitem[{Fjeldbo et~al.(1975)Fjeldbo, Seiden, Sweetnam \&
  Howard}]{Fjeldbo1975}
\bibinfo{author}{Fjeldbo, G.}, \bibinfo{author}{Seiden, B.},
  \bibinfo{author}{Sweetnam, D.}, \& \bibinfo{author}{Howard, T.}
  (\bibinfo{year}{1975}).
\newblock \bibinfo{title}{{The Mariner 10 Radio Occultation Measurements of the
  Ionosphere of Venus}}.
\newblock {\it \bibinfo{journal}{Journal of the Atmospheric Sciences}\/},  {\it
  \bibinfo{volume}{32}\/}\bibinfo{issue}{(6)}, \bibinfo{pages}{1232--1236}.
  \URLprefix
  \url{http://journals.ametsoc.org/doi/abs/10.1175/1520-0469{\%}281975{\%}29032{\%}3C1232{\%}3ATMROMO{\%}3E2.0.CO{\%}3B2}.
  \DOIprefix\doi{10.1175/1520-0469(1975)032<1232:TMROMO>2.0.CO;2}.
\bibitem[{Gavrik et~al.(2009)Gavrik, Pavelyev \& Gavrik}]{Gavrik2009}
\bibinfo{author}{Gavrik, A.~L.}, \bibinfo{author}{Pavelyev, A.~G.}, \&
  \bibinfo{author}{Gavrik, Y.~A.} (\bibinfo{year}{2009}).
\newblock \bibinfo{title}{{Detection of ionospheric layers in the dayside
  ionosphere of Venus at altitudes of 80-120 km from Venera-15 and -16
  two-frequency radio-occultation results}}.
\newblock {\it \bibinfo{journal}{Geomagnetism and Aeronomy}\/},  {\it
  \bibinfo{volume}{49}\/}\bibinfo{issue}{(8)}, \bibinfo{pages}{1223--1225}.
  \DOIprefix\doi{10.1134/S0016793209080362}.
\bibitem[{G{\'{e}}rard et~al.(2017)G{\'{e}}rard, Bougher, L{\'{o}}pez-Valverde,
  P{\"{a}}tzold, Drossart \& Piccioni}]{Gerard2017}
\bibinfo{author}{G{\'{e}}rard, J.-C.}, \bibinfo{author}{Bougher, S.~W.},
  \bibinfo{author}{L{\'{o}}pez-Valverde, M.~A.},
  \bibinfo{author}{P{\"{a}}tzold, M.}, \bibinfo{author}{Drossart, P.}, \&
  \bibinfo{author}{Piccioni, G.} (\bibinfo{year}{2017}).
\newblock \bibinfo{title}{{Aeronomy of the Venus Upper Atmosphere}}.
\newblock {\it \bibinfo{journal}{Space Science Reviews}\/},  {\it
  \bibinfo{volume}{212}\/}\bibinfo{issue}{(3)}, \bibinfo{pages}{1617--1683}.
  \URLprefix \url{https://doi.org/10.1007/s11214-017-0422-0}.
  \DOIprefix\doi{10.1007/s11214-017-0422-0}.
\bibitem[{{Gunnar Fjeldbo} \& Eshleman(1966)}]{Fjeldbo1966Mariner}
\bibinfo{author}{{Gunnar Fjeldbo}, W. C.~F.}, \& \bibinfo{author}{Eshleman,
  V.~R.} (\bibinfo{year}{1966}).
\newblock \bibinfo{title}{{Atmosphere of Mars: Mariner IV Models Compared}}, .
\newblock {\it \bibinfo{volume}{16}\/}\bibinfo{issue}{(1926)},
  \bibinfo{pages}{812--815}.
\bibitem[{H{\"{a}}usler et~al.(2006)H{\"{a}}usler, P{\"{a}}tzold, Tyler,
  Simpson, Bird, Dehant, Barriot, Eidel, Mattei, Remus, Selle, Tellmann \&
  Imamura}]{Hausler2006}
\bibinfo{author}{H{\"{a}}usler, B.}, \bibinfo{author}{P{\"{a}}tzold, M.},
  \bibinfo{author}{Tyler, G.~L.}, \bibinfo{author}{Simpson, R.~A.},
  \bibinfo{author}{Bird, M.~K.}, \bibinfo{author}{Dehant, V.},
  \bibinfo{author}{Barriot, J.~P.}, \bibinfo{author}{Eidel, W.},
  \bibinfo{author}{Mattei, R.}, \bibinfo{author}{Remus, S.},
  \bibinfo{author}{Selle, J.}, \bibinfo{author}{Tellmann, S.}, \&
  \bibinfo{author}{Imamura, T.} (\bibinfo{year}{2006}).
\newblock \bibinfo{title}{{Radio science investigations by VeRa onboard the
  Venus Express spacecraft}}.
\newblock {\it \bibinfo{journal}{Planetary and Space Science}\/},  {\it
  \bibinfo{volume}{54}\/}\bibinfo{issue}{(13-14)}, \bibinfo{pages}{1315--1335}.
  \DOIprefix\doi{10.1016/j.pss.2006.04.032}.
\bibitem[{Hensley et~al.(2020)Hensley, Withers, Girazian, P{\"{a}}tzold,
  Tellmann \& H{\"{a}}usler}]{Hensley2020}
\bibinfo{author}{Hensley, K.}, \bibinfo{author}{Withers, P.},
  \bibinfo{author}{Girazian, Z.}, \bibinfo{author}{P{\"{a}}tzold, M.},
  \bibinfo{author}{Tellmann, S.}, \& \bibinfo{author}{H{\"{a}}usler, B.}
  (\bibinfo{year}{2020}).
\newblock \bibinfo{title}{{Dependence of Dayside Electron Densities at Venus on
  Solar Irradiance}}.
\newblock {\it \bibinfo{journal}{Journal of Geophysical Research: Space
  Physics}\/},  {\it \bibinfo{volume}{125}\/}\bibinfo{issue}{(2)}.
  \DOIprefix\doi{10.1029/2019ja027167}.
\bibitem[{Hinson \& Jenkins(1995)}]{Hinson1995}
\bibinfo{author}{Hinson, D.~P.}, \& \bibinfo{author}{Jenkins, J.~M.}
  (\bibinfo{year}{1995}).
\newblock \bibinfo{title}{{Magellan radio occultation measurements of
  atmospheric waves on venus}}.
\newblock \DOIprefix\doi{10.1006/icar.1995.1064}.
\bibitem[{Hueso et~al.(2012)Hueso, Peralta \& S{\'{a}}nchez-Lavega}]{Hueso2012}
\bibinfo{author}{Hueso, R.}, \bibinfo{author}{Peralta, J.}, \&
  \bibinfo{author}{S{\'{a}}nchez-Lavega, A.} (\bibinfo{year}{2012}).
\newblock \bibinfo{title}{{Assessing the long-term variability of Venus winds
  at cloud level from VIRTIS-Venus Express}}.
\newblock {\it \bibinfo{journal}{Icarus}\/},  {\it
  \bibinfo{volume}{217}\/}\bibinfo{issue}{(2)}, \bibinfo{pages}{585--598}.
  \URLprefix \url{http://dx.doi.org/10.1016/j.icarus.2011.04.020}.
  \DOIprefix\doi{10.1016/j.icarus.2011.04.020}.
\bibitem[{Imamura et~al.(2017)Imamura, Ando, Tellmann, P{\"{a}}tzold,
  H{\"{a}}usler, Yamazaki, Sato, Noguchi, Futaana, Oschlisniok, Limaye,
  Choudhary, Murata, Takeuchi, Hirose, Ichikawa, Toda, Tomiki, Abe, Yamamoto,
  Noda, Iwata, Murakami, Satoh, Fukuhara, Ogohara, Sugiyama, Kashimura,
  Ohtsuki, Takagi, Yamamoto, Hirata, Hashimoto, Yamada, Suzuki, Ishii,
  Hayashiyama, Lee \& Nakamura}]{Imamura2017}
\bibinfo{author}{Imamura, T.}, \bibinfo{author}{Ando, H.},
  \bibinfo{author}{Tellmann, S.}, \bibinfo{author}{P{\"{a}}tzold, M.},
  \bibinfo{author}{H{\"{a}}usler, B.}, \bibinfo{author}{Yamazaki, A.},
  \bibinfo{author}{Sato, T.~M.}, \bibinfo{author}{Noguchi, K.},
  \bibinfo{author}{Futaana, Y.}, \bibinfo{author}{Oschlisniok, J.},
  \bibinfo{author}{Limaye, S.}, \bibinfo{author}{Choudhary, R.~K.},
  \bibinfo{author}{Murata, Y.}, \bibinfo{author}{Takeuchi, H.},
  \bibinfo{author}{Hirose, C.}, \bibinfo{author}{Ichikawa, T.},
  \bibinfo{author}{Toda, T.}, \bibinfo{author}{Tomiki, A.},
  \bibinfo{author}{Abe, T.}, \bibinfo{author}{Yamamoto, Z.~I.},
  \bibinfo{author}{Noda, H.}, \bibinfo{author}{Iwata, T.},
  \bibinfo{author}{Murakami, S.~Y.}, \bibinfo{author}{Satoh, T.},
  \bibinfo{author}{Fukuhara, T.}, \bibinfo{author}{Ogohara, K.},
  \bibinfo{author}{Sugiyama, K.~I.}, \bibinfo{author}{Kashimura, H.},
  \bibinfo{author}{Ohtsuki, S.}, \bibinfo{author}{Takagi, S.},
  \bibinfo{author}{Yamamoto, Y.}, \bibinfo{author}{Hirata, N.},
  \bibinfo{author}{Hashimoto, G.~L.}, \bibinfo{author}{Yamada, M.},
  \bibinfo{author}{Suzuki, M.}, \bibinfo{author}{Ishii, N.},
  \bibinfo{author}{Hayashiyama, T.}, \bibinfo{author}{Lee, Y.~J.}, \&
  \bibinfo{author}{Nakamura, M.} (\bibinfo{year}{2017}).
\newblock \bibinfo{title}{{Initial performance of the radio occultation
  experiment in the Venus orbiter mission Akatsuki Akatsuki at Venus: The First
  Year of Scientific Operation Masato Nakamura, Dmitri Titov, Kevin
  McGouldrick, Pierre Drossart, Jean-Loup Bertaux and Huixin Liu 7. }}.
\newblock {\it \bibinfo{journal}{Earth, Planets and Space}\/},  {\it
  \bibinfo{volume}{69}\/}\bibinfo{issue}{(1)}.
  \DOIprefix\doi{10.1186/s40623-017-0722-3}.
\bibitem[{Imamura et~al.(2018)Imamura, Miyamoto, Ando, H{\"a}usler,
  P{\"a}tzold, Tellmann, Tsuda, Aoyama, Murata, Takeuchi
  et~al.}]{imamura2018fine}
\bibinfo{author}{Imamura, T.}, \bibinfo{author}{Miyamoto, M.},
  \bibinfo{author}{Ando, H.}, \bibinfo{author}{H{\"a}usler, B.},
  \bibinfo{author}{P{\"a}tzold, M.}, \bibinfo{author}{Tellmann, S.},
  \bibinfo{author}{Tsuda, T.}, \bibinfo{author}{Aoyama, Y.},
  \bibinfo{author}{Murata, Y.}, \bibinfo{author}{Takeuchi, H.} et~al.
  (\bibinfo{year}{2018}).
\newblock \bibinfo{title}{Fine vertical structures at the cloud heights of
  venus revealed by radio holographic analysis of venus express and akatsuki
  radio occultation data}.
\newblock {\it \bibinfo{journal}{Journal of Geophysical Research: Planets}\/},
  {\it \bibinfo{volume}{123}\/}\bibinfo{issue}{(8)},
  \bibinfo{pages}{2151--2161}.
\bibitem[{Imamura et~al.(2011)Imamura, Toda, Tomiki, Hirahara, Hayashiyama,
  Mochizuki, Yamamoto, Abe, Iwata, Noda, Futaana, Ando, H{\"{a}}usler,
  P{\"{a}}tzold \& Nabatov}]{Imamura2011}
\bibinfo{author}{Imamura, T.}, \bibinfo{author}{Toda, T.},
  \bibinfo{author}{Tomiki, A.}, \bibinfo{author}{Hirahara, D.},
  \bibinfo{author}{Hayashiyama, T.}, \bibinfo{author}{Mochizuki, N.},
  \bibinfo{author}{Yamamoto, Z.~I.}, \bibinfo{author}{Abe, T.},
  \bibinfo{author}{Iwata, T.}, \bibinfo{author}{Noda, H.},
  \bibinfo{author}{Futaana, Y.}, \bibinfo{author}{Ando, H.},
  \bibinfo{author}{H{\"{a}}usler, B.}, \bibinfo{author}{P{\"{a}}tzold, M.}, \&
  \bibinfo{author}{Nabatov, A.} (\bibinfo{year}{2011}).
\newblock \bibinfo{title}{{Radio occultation experiment of the Venus atmosphere
  and ionosphere with the Venus orbiter Akatsuki}}.
\newblock {\it \bibinfo{journal}{Earth, Planets and Space}\/},  {\it
  \bibinfo{volume}{63}\/}\bibinfo{issue}{(6)}, \bibinfo{pages}{493--501}.
  \DOIprefix\doi{10.5047/eps.2011.03.009}.
\bibitem[{Ivanov-Kholodny et~al.(1979)Ivanov-Kholodny, Kolosov, Savich,
  Alexandrov, Vasilyev, Vyshlov, Dubrovin, Zaitsev, Michailov, Petrov, Samovol,
  Samoznaev, Sidorenko \& Hasyanov}]{Ivanov-Kholodny1979}
\bibinfo{author}{Ivanov-Kholodny, G.~S.}, \bibinfo{author}{Kolosov, M.~A.},
  \bibinfo{author}{Savich, N.~A.}, \bibinfo{author}{Alexandrov, Y.~N.},
  \bibinfo{author}{Vasilyev, M.~B.}, \bibinfo{author}{Vyshlov, A.~S.},
  \bibinfo{author}{Dubrovin, V.~M.}, \bibinfo{author}{Zaitsev, A.~L.},
  \bibinfo{author}{Michailov, A.~V.}, \bibinfo{author}{Petrov, G.~M.},
  \bibinfo{author}{Samovol, V.~A.}, \bibinfo{author}{Samoznaev, L.~N.},
  \bibinfo{author}{Sidorenko, A.~I.}, \& \bibinfo{author}{Hasyanov, A.~F.}
  (\bibinfo{year}{1979}).
\newblock \bibinfo{title}{{Daytime ionosphere of Venus as studied with Veneras
  9 and 10 dual-frequency radio occultation experiments}}.
\newblock {\it \bibinfo{journal}{Icarus}\/},  {\it
  \bibinfo{volume}{39}\/}\bibinfo{issue}{(2)}, \bibinfo{pages}{209--213}.
  \DOIprefix\doi{10.1016/0019-1035(79)90164-7}.
\bibitem[{Jenkins \& Hinson(1997)}]{1997DPS}
\bibinfo{author}{Jenkins, J.}, \& \bibinfo{author}{Hinson, D.}
  (\bibinfo{year}{1997}).
\newblock \bibinfo{title}{{Magellan Radio Occultation Studies of Venus'
  Atmosphere (1991--1994)}}.
\newblock In {\it \bibinfo{booktitle}{AAS/Division for Planetary Sciences
  Meeting Abstracts {\#}29}\/} AAS/Division for Planetary Sciences Meeting
  Abstracts (p. \bibinfo{pages}{35.10}).
\bibitem[{Jenkins et~al.(1994)Jenkins, Steffes, Hinson, Twicken \&
  Tyler}]{Jenkins1994}
\bibinfo{author}{Jenkins, J.~M.}, \bibinfo{author}{Steffes, P.~G.},
  \bibinfo{author}{Hinson, D.~P.}, \bibinfo{author}{Twicken, J.~D.}, \&
  \bibinfo{author}{Tyler, G.~L.} (\bibinfo{year}{1994}).
\newblock \bibinfo{title}{{Radio Occultation Studies of the Venus Atmosphere
  with the Magellan Spacecraft. 2. Results from the October 1991 Experiments}}.
\newblock \DOIprefix\doi{10.1006/icar.1994.1108}.
\bibitem[{Kliore et~al.(1965)Kliore, Cain, Levy, Eshleman, Fjeldbo \&
  Drake}]{Kliore1965}
\bibinfo{author}{Kliore, A.}, \bibinfo{author}{Cain, D.~L.},
  \bibinfo{author}{Levy, G.~S.}, \bibinfo{author}{Eshleman, V.~R.},
  \bibinfo{author}{Fjeldbo, G.}, \& \bibinfo{author}{Drake, F.~D.}
  (\bibinfo{year}{1965}).
\newblock \bibinfo{title}{{Occultation Experiment: Results of the First Direct
  Measurement of Mars's Atmosphere and Ionosphere}}.
\newblock {\it \bibinfo{journal}{Science}\/},  {\it
  \bibinfo{volume}{149}\/}\bibinfo{issue}{(3689)}, \bibinfo{pages}{1243--1248}.
  \URLprefix
  \url{https://www.sciencemag.org/lookup/doi/10.1126/science.149.3689.1243}.
  \DOIprefix\doi{10.1126/science.149.3689.1243}.
\bibitem[{Kliore et~al.(1967)Kliore, Levy, Cain, Fjeldbo \&
  Rasool}]{Kliore1967}
\bibinfo{author}{Kliore, A.}, \bibinfo{author}{Levy, G.~S.},
  \bibinfo{author}{Cain, D.~L.}, \bibinfo{author}{Fjeldbo, G.}, \&
  \bibinfo{author}{Rasool, S.~I.} (\bibinfo{year}{1967}).
\newblock \bibinfo{title}{{Atmosphere and ionosphere of venus from the Mariner
  V S-band radio occultation measurement}}.
\newblock {\it \bibinfo{journal}{Science}\/},  {\it
  \bibinfo{volume}{158}\/}\bibinfo{issue}{(3809)}, \bibinfo{pages}{1683--1688}.
  \DOIprefix\doi{10.1126/science.158.3809.1683}.
\bibitem[{Kliore \& Patel(1980)}]{Kliore1980}
\bibinfo{author}{Kliore, A.~J.}, \& \bibinfo{author}{Patel, I.~R.}
  (\bibinfo{year}{1980}).
\newblock \bibinfo{title}{{Vertical structure of the atmosphere of Venus from
  Pioneer Venus Orbiter radio occultations}}.
\newblock \DOIprefix\doi{10.1029/ja085ia13p07957}.
\bibitem[{Kliore et~al.(1979)Kliore, Patel, Nagy, Cravens \&
  Gombosi}]{Kliore1979}
\bibinfo{author}{Kliore, A.~J.}, \bibinfo{author}{Patel, I.~R.},
  \bibinfo{author}{Nagy, A.~F.}, \bibinfo{author}{Cravens, T.~E.}, \&
  \bibinfo{author}{Gombosi, T.~I.} (\bibinfo{year}{1979}).
\newblock \bibinfo{title}{{Initial observations of the nightside ionosphere of
  Venus from Pioneer Venus orbiter radio occultations}}.
\newblock {\it \bibinfo{journal}{Science}\/},  {\it
  \bibinfo{volume}{205}\/}\bibinfo{issue}{(4401)}, \bibinfo{pages}{99--102}.
  \DOIprefix\doi{10.1126/science.205.4401.99}.
\bibitem[{Kolosov et~al.(1979)Kolosov, Yakovlev, Efimov, Pavelyev \&
  Matyugov}]{Kolosov1979}
\bibinfo{author}{Kolosov, M.~A.}, \bibinfo{author}{Yakovlev, O.~I.},
  \bibinfo{author}{Efimov, A.~I.}, \bibinfo{author}{Pavelyev, A.~G.}, \&
  \bibinfo{author}{Matyugov, S.~S.} (\bibinfo{year}{1979}).
\newblock \bibinfo{title}{{Radio occultation of the Venusian atmosphere and
  bistatic radiolocation of the surface of Venus using the Venera‐9 and
  Venera‐10 satellites}}.
\newblock \DOIprefix\doi{10.1029/RS014i001p00163}.
\bibitem[{Konopliv et~al.(1999)Konopliv, Banerdt \& Sjogren}]{Konopliv1999}
\bibinfo{author}{Konopliv, A.~S.}, \bibinfo{author}{Banerdt, W.~B.}, \&
  \bibinfo{author}{Sjogren, W.~L.} (\bibinfo{year}{1999}).
\newblock \bibinfo{title}{{Venus Gravity: 180th Degree and Order Model}}.
\newblock {\it \bibinfo{journal}{Icarus}\/},  {\it
  \bibinfo{volume}{139}\/}\bibinfo{issue}{(1)}, \bibinfo{pages}{3--18}.
  \DOIprefix\doi{10.1006/icar.1999.6086}.
\bibitem[{Kursinski et~al.(2000)Kursinski, Hajj, Leroy \&
  Herman}]{Kursinski2000}
\bibinfo{author}{Kursinski, E.}, \bibinfo{author}{Hajj, G.-A.},
  \bibinfo{author}{Leroy, S.}, \& \bibinfo{author}{Herman, B.}
  (\bibinfo{year}{2000}).
\newblock \bibinfo{title}{{The GPS radio Occulation techniques}}.
\newblock \URLprefix \url{http://hdl.handle.net/2014/14027}.
\bibitem[{Lee et~al.(2012)Lee, Titov, Tellmann, Piccialli, Ignatiev,
  P{\"{a}}tzold, H{\"{a}}usler, Piccioni \& Drossart}]{Lee2012}
\bibinfo{author}{Lee, Y.~J.}, \bibinfo{author}{Titov, D.~V.},
  \bibinfo{author}{Tellmann, S.}, \bibinfo{author}{Piccialli, A.},
  \bibinfo{author}{Ignatiev, N.}, \bibinfo{author}{P{\"{a}}tzold, M.},
  \bibinfo{author}{H{\"{a}}usler, B.}, \bibinfo{author}{Piccioni, G.}, \&
  \bibinfo{author}{Drossart, P.} (\bibinfo{year}{2012}).
\newblock \bibinfo{title}{{Vertical structure of the Venus cloud top from the
  VeRa and VIRTIS observations onboard Venus Express}}.
\newblock {\it \bibinfo{journal}{Icarus}\/},  {\it
  \bibinfo{volume}{217}\/}\bibinfo{issue}{(2)}, \bibinfo{pages}{599--609}.
  \URLprefix \url{http://dx.doi.org/10.1016/j.icarus.2011.07.001}.
  \DOIprefix\doi{10.1016/j.icarus.2011.07.001}.
\bibitem[{Limaye et~al.(2017)Limaye, Lebonnois, Mahieux, P{\"{a}}tzold,
  Bougher, Bruinsma, Chamberlain, Clancy, G{\'{e}}rard, Gilli, Grassi, Haus,
  Herrmann, Imamura, Kohler, Krause, Migliorini, Montmessin, Pere, Persson,
  Piccialli, Rengel, Rodin, Sandor, Sornig, Svedhem, Tellmann, Tanga, Vandaele,
  Widemann, Wilson, M{\"{u}}ller-Wodarg \& Zasova}]{Limaye2017}
\bibinfo{author}{Limaye, S.~S.}, \bibinfo{author}{Lebonnois, S.},
  \bibinfo{author}{Mahieux, A.}, \bibinfo{author}{P{\"{a}}tzold, M.},
  \bibinfo{author}{Bougher, S.}, \bibinfo{author}{Bruinsma, S.},
  \bibinfo{author}{Chamberlain, S.}, \bibinfo{author}{Clancy, R.~T.},
  \bibinfo{author}{G{\'{e}}rard, J.~C.}, \bibinfo{author}{Gilli, G.},
  \bibinfo{author}{Grassi, D.}, \bibinfo{author}{Haus, R.},
  \bibinfo{author}{Herrmann, M.}, \bibinfo{author}{Imamura, T.},
  \bibinfo{author}{Kohler, E.}, \bibinfo{author}{Krause, P.},
  \bibinfo{author}{Migliorini, A.}, \bibinfo{author}{Montmessin, F.},
  \bibinfo{author}{Pere, C.}, \bibinfo{author}{Persson, M.},
  \bibinfo{author}{Piccialli, A.}, \bibinfo{author}{Rengel, M.},
  \bibinfo{author}{Rodin, A.}, \bibinfo{author}{Sandor, B.},
  \bibinfo{author}{Sornig, M.}, \bibinfo{author}{Svedhem, H.},
  \bibinfo{author}{Tellmann, S.}, \bibinfo{author}{Tanga, P.},
  \bibinfo{author}{Vandaele, A.~C.}, \bibinfo{author}{Widemann, T.},
  \bibinfo{author}{Wilson, C.~F.}, \bibinfo{author}{M{\"{u}}ller-Wodarg, I.},
  \& \bibinfo{author}{Zasova, L.} (\bibinfo{year}{2017}).
\newblock \bibinfo{title}{{The thermal structure of the Venus atmosphere:
  Intercomparison of Venus Express and ground based observations of vertical
  temperature and density profiles}}.
\newblock {\it \bibinfo{journal}{Icarus}\/},  {\it \bibinfo{volume}{294}\/},
  \bibinfo{pages}{124--155}. \DOIprefix\doi{10.1016/j.icarus.2017.04.020}.
\bibitem[{NASA()}]{nasa}
\bibinfo{author}{NASA} ().
\newblock \bibinfo{title}{{Venus fact sheet,
  https://nssdc.gsfc.nasa.gov/planetary/factsheet/venusfact.html}}.
\newblock \URLprefix
  \url{https://nssdc.gsfc.nasa.gov/planetary/factsheet/venusfact.html}.
\bibitem[{Newman et~al.(1984)Newman, Schubert, Kliore \& Patel}]{Newman1984}
\bibinfo{author}{Newman, M.}, \bibinfo{author}{Schubert, G.},
  \bibinfo{author}{Kliore, A.~J.}, \& \bibinfo{author}{Patel, I.~R.}
  (\bibinfo{year}{1984}).
\newblock \bibinfo{title}{{Zonal Winds in the Middle Atmosphere of Venus From
  Pioneer Venus Radio Occultation Data.}}
\newblock \DOIprefix\doi{10.1175/1520-0469(1984)041<1901:ZWITMA>2.0.CO;2}.
\bibitem[{Paik \& Asmar(2011)}]{Paik2011}
\bibinfo{author}{Paik, M.}, \& \bibinfo{author}{Asmar, S.~W.}
  (\bibinfo{year}{2011}).
\newblock \bibinfo{title}{{Detecting high dynamics signals from open-loop radio
  science investigations}}.
\newblock {\it \bibinfo{journal}{Proceedings of the IEEE}\/},  {\it
  \bibinfo{volume}{99}\/}\bibinfo{issue}{(5)}, \bibinfo{pages}{881--888}.
  \DOIprefix\doi{10.1109/JPROC.2010.2084550}.
\bibitem[{{Park} et~al.(2021){Park}, {Folkner}, {Williams} \&
  {Boggs}}]{2021AJ....161..105P}
\bibinfo{author}{{Park}, R.~S.}, \bibinfo{author}{{Folkner}, W.~M.},
  \bibinfo{author}{{Williams}, J.~G.}, \& \bibinfo{author}{{Boggs}, D.~H.}
  (\bibinfo{year}{2021}).
\newblock \bibinfo{title}{{The JPL Planetary and Lunar Ephemerides DE440 and
  DE441}}.
\newblock {\it \bibinfo{journal}{The Astronomical Journal}\/},  {\it
  \bibinfo{volume}{161}\/}\bibinfo{issue}{(3)}, \bibinfo{pages}{105}.
  \DOIprefix\doi{10.3847/1538-3881/abd414}.
\bibitem[{P{\"{a}}tzold et~al.(2007)P{\"{a}}tzold, H{\"{a}}usler, Bird,
  Tellmann, Mattei, Asmar, Dehant, Eidel, Imamura, Simpson \&
  Tyler}]{Patzold2007}
\bibinfo{author}{P{\"{a}}tzold, M.}, \bibinfo{author}{H{\"{a}}usler, B.},
  \bibinfo{author}{Bird, M.~K.}, \bibinfo{author}{Tellmann, S.},
  \bibinfo{author}{Mattei, R.}, \bibinfo{author}{Asmar, S.~W.},
  \bibinfo{author}{Dehant, V.}, \bibinfo{author}{Eidel, W.},
  \bibinfo{author}{Imamura, T.}, \bibinfo{author}{Simpson, R.~A.}, \&
  \bibinfo{author}{Tyler, G.~L.} (\bibinfo{year}{2007}).
\newblock \bibinfo{title}{{The structure of Venus' middle atmosphere and
  ionosphere}}.
\newblock {\it \bibinfo{journal}{Nature}\/},  {\it
  \bibinfo{volume}{450}\/}\bibinfo{issue}{(7170)}, \bibinfo{pages}{657--660}.
  \DOIprefix\doi{10.1038/nature06239}.
\bibitem[{Peter et~al.(2014)Peter, P{\"{a}}tzold, Molina-Cuberos, Witasse,
  Gonz{\'{a}}lez-Galindo, Withers, Bird, H{\"{a}}usler, Hinson, Tellmann \&
  Tyler}]{Peter2014}
\bibinfo{author}{Peter, K.}, \bibinfo{author}{P{\"{a}}tzold, M.},
  \bibinfo{author}{Molina-Cuberos, G.}, \bibinfo{author}{Witasse, O.},
  \bibinfo{author}{Gonz{\'{a}}lez-Galindo, F.}, \bibinfo{author}{Withers, P.},
  \bibinfo{author}{Bird, M.~K.}, \bibinfo{author}{H{\"{a}}usler, B.},
  \bibinfo{author}{Hinson, D.~P.}, \bibinfo{author}{Tellmann, S.}, \&
  \bibinfo{author}{Tyler, G.~L.} (\bibinfo{year}{2014}).
\newblock \bibinfo{title}{{The dayside ionospheres of Mars and Venus: Comparing
  a one-dimensional photochemical model with MaRS (Mars Express) and VeRa
  (Venus Express) observations}}.
\newblock {\it \bibinfo{journal}{Icarus}\/},  {\it \bibinfo{volume}{233}\/},
  \bibinfo{pages}{66--82}. \URLprefix
  \url{http://dx.doi.org/10.1016/j.icarus.2014.01.028}.
  \DOIprefix\doi{10.1016/j.icarus.2014.01.028}.
\bibitem[{Phinney \& Anderson(1968)}]{Phinney1968}
\bibinfo{author}{Phinney, R.~A.}, \& \bibinfo{author}{Anderson, D.~L.}
  (\bibinfo{year}{1968}).
\newblock \bibinfo{title}{{On the radio occultation method for studying
  planetary atmospheres}}.
\newblock {\it \bibinfo{journal}{Journal of Geophysical Research}\/},  {\it
  \bibinfo{volume}{73}\/}\bibinfo{issue}{(5)}, \bibinfo{pages}{1819--1827}.
  \DOIprefix\doi{10.1029/ja073i005p01819}.
\bibitem[{Piccialli et~al.(2012)Piccialli, Tellmann, Titov, Limaye, Khatuntsev,
  P{\"{a}}tzold \& H{\"{a}}usler}]{Piccialli2012}
\bibinfo{author}{Piccialli, A.}, \bibinfo{author}{Tellmann, S.},
  \bibinfo{author}{Titov, D.~V.}, \bibinfo{author}{Limaye, S.~S.},
  \bibinfo{author}{Khatuntsev, I.~V.}, \bibinfo{author}{P{\"{a}}tzold, M.}, \&
  \bibinfo{author}{H{\"{a}}usler, B.} (\bibinfo{year}{2012}).
\newblock \bibinfo{title}{{Dynamical properties of the Venus mesosphere from
  the radio-occultation experiment VeRa onboard Venus Express}}.
\newblock {\it \bibinfo{journal}{Icarus}\/},  {\it
  \bibinfo{volume}{217}\/}\bibinfo{issue}{(2)}, \bibinfo{pages}{669--681}.
  \URLprefix \url{http://dx.doi.org/10.1016/j.icarus.2011.07.016}.
  \DOIprefix\doi{10.1016/j.icarus.2011.07.016}.
\bibitem[{Schinder et~al.(2015)Schinder, Flasar, Marouf, French, Anabtawi,
  Barbinis \& Kliore}]{Schinder2015}
\bibinfo{author}{Schinder, P.~J.}, \bibinfo{author}{Flasar, F.~M.},
  \bibinfo{author}{Marouf, E.~A.}, \bibinfo{author}{French, R.~G.},
  \bibinfo{author}{Anabtawi, A.}, \bibinfo{author}{Barbinis, E.}, \&
  \bibinfo{author}{Kliore, A.~J.} (\bibinfo{year}{2015}).
\newblock \bibinfo{title}{{A numerical technique for two-way radio occultations
  by oblate axisymmetric atmospheres with zonal winds}}.
\newblock {\it \bibinfo{journal}{Radio Science}\/},  {\it
  \bibinfo{volume}{50}\/}\bibinfo{issue}{(7)}, \bibinfo{pages}{712--727}.
  \DOIprefix\doi{10.1002/2015RS005690}.
\bibitem[{Schinder et~al.(2011)Schinder, Flasar, Marouf, French, McGhee,
  Kliore, Rappaport, Barbinis, Fleischman \& Anabtawi}]{Schinder2011}
\bibinfo{author}{Schinder, P.~J.}, \bibinfo{author}{Flasar, F.~M.},
  \bibinfo{author}{Marouf, E.~A.}, \bibinfo{author}{French, R.~G.},
  \bibinfo{author}{McGhee, C.~A.}, \bibinfo{author}{Kliore, A.~J.},
  \bibinfo{author}{Rappaport, N.~J.}, \bibinfo{author}{Barbinis, E.},
  \bibinfo{author}{Fleischman, D.}, \& \bibinfo{author}{Anabtawi, A.}
  (\bibinfo{year}{2011}).
\newblock \bibinfo{title}{{The structure of Titan's atmosphere from Cassini
  radio occultations}}.
\newblock {\it \bibinfo{journal}{Icarus}\/},  {\it
  \bibinfo{volume}{215}\/}\bibinfo{issue}{(2)}, \bibinfo{pages}{460--474}.
  \URLprefix \url{http://dx.doi.org/10.1016/j.icarus.2011.07.030}.
  \DOIprefix\doi{10.1016/j.icarus.2011.07.030}.
\bibitem[{Schinder et~al.(2012)Schinder, Flasar, Marouf, French, McGhee,
  Kliore, Rappaport, Barbinis, Fleischman \& Anabtawi}]{Schinder2012}
\bibinfo{author}{Schinder, P.~J.}, \bibinfo{author}{Flasar, F.~M.},
  \bibinfo{author}{Marouf, E.~A.}, \bibinfo{author}{French, R.~G.},
  \bibinfo{author}{McGhee, C.~A.}, \bibinfo{author}{Kliore, A.~J.},
  \bibinfo{author}{Rappaport, N.~J.}, \bibinfo{author}{Barbinis, E.},
  \bibinfo{author}{Fleischman, D.}, \& \bibinfo{author}{Anabtawi, A.}
  (\bibinfo{year}{2012}).
\newblock \bibinfo{title}{{The structure of Titan's atmosphere from Cassini
  radio occultations: Occultations from the Prime and Equinox missions}}.
\newblock {\it \bibinfo{journal}{Icarus}\/},  {\it
  \bibinfo{volume}{221}\/}\bibinfo{issue}{(2)}, \bibinfo{pages}{1020--1031}.
  \URLprefix \url{http://dx.doi.org/10.1016/j.icarus.2012.10.021}.
  \DOIprefix\doi{10.1016/j.icarus.2012.10.021}.
\bibitem[{Svedhem et~al.(2007)Svedhem, Titov, McCoy, Lebreton, Barabash,
  Bertaux, Drossart, Formisano, H{\"{a}}usler, Korablev, Markiewicz, Nevejans,
  P{\"{a}}tzold, Piccioni, Zhang, Taylor, Lellouch, Koschny, Witasse, Eggel,
  Warhaut, Accomazzo, Rodriguez-Canabal, Fabrega, Schirmann, Clochet \&
  Coradini}]{Svedhem2007}
\bibinfo{author}{Svedhem, H.}, \bibinfo{author}{Titov, D.~V.},
  \bibinfo{author}{McCoy, D.}, \bibinfo{author}{Lebreton, J.~P.},
  \bibinfo{author}{Barabash, S.}, \bibinfo{author}{Bertaux, J.~L.},
  \bibinfo{author}{Drossart, P.}, \bibinfo{author}{Formisano, V.},
  \bibinfo{author}{H{\"{a}}usler, B.}, \bibinfo{author}{Korablev, O.},
  \bibinfo{author}{Markiewicz, W.~J.}, \bibinfo{author}{Nevejans, D.},
  \bibinfo{author}{P{\"{a}}tzold, M.}, \bibinfo{author}{Piccioni, G.},
  \bibinfo{author}{Zhang, T.~L.}, \bibinfo{author}{Taylor, F.~W.},
  \bibinfo{author}{Lellouch, E.}, \bibinfo{author}{Koschny, D.},
  \bibinfo{author}{Witasse, O.}, \bibinfo{author}{Eggel, H.},
  \bibinfo{author}{Warhaut, M.}, \bibinfo{author}{Accomazzo, A.},
  \bibinfo{author}{Rodriguez-Canabal, J.}, \bibinfo{author}{Fabrega, J.},
  \bibinfo{author}{Schirmann, T.}, \bibinfo{author}{Clochet, A.}, \&
  \bibinfo{author}{Coradini, M.} (\bibinfo{year}{2007}).
\newblock \bibinfo{title}{{Venus Express-The first European mission to Venus}}.
\newblock {\it \bibinfo{journal}{Planetary and Space Science}\/},  {\it
  \bibinfo{volume}{55}\/}\bibinfo{issue}{(12)}, \bibinfo{pages}{1636--1652}.
  \DOIprefix\doi{10.1016/j.pss.2007.01.013}.
\bibitem[{Tellmann et~al.(2012)Tellmann, H{\"{a}}usler, Hinson, Tyler, Andert,
  Bird, Imamura, P{\"{a}}tzold \& Remus}]{Tellmann2012}
\bibinfo{author}{Tellmann, S.}, \bibinfo{author}{H{\"{a}}usler, B.},
  \bibinfo{author}{Hinson, D.~P.}, \bibinfo{author}{Tyler, G.~L.},
  \bibinfo{author}{Andert, T.~P.}, \bibinfo{author}{Bird, M.~K.},
  \bibinfo{author}{Imamura, T.}, \bibinfo{author}{P{\"{a}}tzold, M.}, \&
  \bibinfo{author}{Remus, S.} (\bibinfo{year}{2012}).
\newblock \bibinfo{title}{{Small-scale temperature fluctuations seen by the
  VeRa Radio Science Experiment on Venus Express}}.
\newblock {\it \bibinfo{journal}{Icarus}\/},  {\it
  \bibinfo{volume}{221}\/}\bibinfo{issue}{(2)}, \bibinfo{pages}{471--480}.
  \URLprefix \url{http://dx.doi.org/10.1016/j.icarus.2012.08.023}.
  \DOIprefix\doi{10.1016/j.icarus.2012.08.023}.
\bibitem[{Tellmann et~al.(2009)Tellmann, P{\"{a}}tzold, H{\"{a}}usler, Bird \&
  {Leonard Tyler}}]{Tellmann2009}
\bibinfo{author}{Tellmann, S.}, \bibinfo{author}{P{\"{a}}tzold, M.},
  \bibinfo{author}{H{\"{a}}usler, B.}, \bibinfo{author}{Bird, M.~K.}, \&
  \bibinfo{author}{{Leonard Tyler}, G.} (\bibinfo{year}{2009}).
\newblock \bibinfo{title}{{Structure of the Venus neutral atmosphere as
  observed by the Radio Science experiment VeRa on Venus Express}}.
\newblock {\it \bibinfo{journal}{Journal of Geophysical Research E:
  Planets}\/},  {\it \bibinfo{volume}{114}\/}\bibinfo{issue}{(4)},
  \bibinfo{pages}{1--19}. \DOIprefix\doi{10.1029/2008JE003204}.
\bibitem[{Titov et~al.(2006)Titov, Svedhem, Koschny, Hoofs, Barabash, Bertaux,
  Drossart, Formisano, H{\"{a}}usler, Korablev, Markiewicz, Nevejans,
  P{\"{a}}tzold, Piccioni, Zhang, Merritt, Witasse, Zender, Accomazzo, Sweeney,
  Trillard, Janvier \& Clochet}]{Titov2006}
\bibinfo{author}{Titov, D.~V.}, \bibinfo{author}{Svedhem, H.},
  \bibinfo{author}{Koschny, D.}, \bibinfo{author}{Hoofs, R.},
  \bibinfo{author}{Barabash, S.}, \bibinfo{author}{Bertaux, J.~L.},
  \bibinfo{author}{Drossart, P.}, \bibinfo{author}{Formisano, V.},
  \bibinfo{author}{H{\"{a}}usler, B.}, \bibinfo{author}{Korablev, O.},
  \bibinfo{author}{Markiewicz, W.~J.}, \bibinfo{author}{Nevejans, D.},
  \bibinfo{author}{P{\"{a}}tzold, M.}, \bibinfo{author}{Piccioni, G.},
  \bibinfo{author}{Zhang, T.~L.}, \bibinfo{author}{Merritt, D.},
  \bibinfo{author}{Witasse, O.}, \bibinfo{author}{Zender, J.},
  \bibinfo{author}{Accomazzo, A.}, \bibinfo{author}{Sweeney, M.},
  \bibinfo{author}{Trillard, D.}, \bibinfo{author}{Janvier, M.}, \&
  \bibinfo{author}{Clochet, A.} (\bibinfo{year}{2006}).
\newblock \bibinfo{title}{{Venus Express science planning}}.
\newblock {\it \bibinfo{journal}{Planetary and Space Science}\/},  {\it
  \bibinfo{volume}{54}\/}\bibinfo{issue}{(13-14)}, \bibinfo{pages}{1279--1297}.
  \DOIprefix\doi{10.1016/j.pss.2006.04.017}.
\bibitem[{Withers(2010)}]{Withers2010}
\bibinfo{author}{Withers, P.} (\bibinfo{year}{2010}).
\newblock \bibinfo{title}{{Prediction of uncertainties in atmospheric
  properties measured by radio occultation experiments}}.
\newblock {\it \bibinfo{journal}{Advances in Space Research}\/},  {\it
  \bibinfo{volume}{46}\/}\bibinfo{issue}{(1)}, \bibinfo{pages}{58--73}.
  \URLprefix \url{http://dx.doi.org/10.1016/j.asr.2010.03.004}.
  \DOIprefix\doi{10.1016/j.asr.2010.03.004}.
\bibitem[{Withers \& Moore(2020)}]{Withers2020}
\bibinfo{author}{Withers, P.}, \& \bibinfo{author}{Moore, L.}
  (\bibinfo{year}{2020}).
\newblock \bibinfo{title}{{How to Process Radio Occultation Data 2 From Time
  Series of Two‐Way Single‐Frequency}}.
\newblock {\it \bibinfo{journal}{Radio Science}\/},  {\it
  \bibinfo{volume}{55}\/}\bibinfo{issue}{(8)}, \bibinfo{pages}{1--25}.
  \DOIprefix\doi{https://doi.org/10.1029/2019RS007046}.
\bibitem[{Withers et~al.(2014)Withers, Moore, Cahoy \& Beerer}]{Withers2014}
\bibinfo{author}{Withers, P.}, \bibinfo{author}{Moore, L.},
  \bibinfo{author}{Cahoy, K.}, \& \bibinfo{author}{Beerer, I.}
  (\bibinfo{year}{2014}).
\newblock \bibinfo{title}{{How to process radio occultation data: 1. From time
  series of frequency residuals to vertical profiles of atmospheric and
  ionospheric properties}}.
\newblock {\it \bibinfo{journal}{Planetary and Space Science}\/},  {\it
  \bibinfo{volume}{101}\/}, \bibinfo{pages}{77--88}. \URLprefix
  \url{http://dx.doi.org/10.1016/j.pss.2014.06.011
  https://linkinghub.elsevier.com/retrieve/pii/S0032063314001822}.
  \DOIprefix\doi{10.1016/j.pss.2014.06.011}.
\bibitem[{Woo(1975)}]{Woo1975}
\bibinfo{author}{Woo, R.} (\bibinfo{year}{1975}).
\newblock \bibinfo{title}{{Observations of Turbulence in the Atmosphere of
  Venus using Mariner 10 Radio Occultation Measurements}}.
\newblock {\it \bibinfo{journal}{Journal of the Atmospheric Sciences}\/},  {\it
  \bibinfo{volume}{32}\/}\bibinfo{issue}{(6)}, \bibinfo{pages}{1084--1090}.
  \URLprefix
  \url{http://journals.ametsoc.org/doi/abs/10.1175/1520-0469{\%}281975{\%}29032{\%}3C1084{\%}3AOOTITA{\%}3E2.0.CO{\%}3B2}.
  \DOIprefix\doi{10.1175/1520-0469(1975)032<1084:OOTITA>2.0.CO;2}.
\bibitem[{Yakovlev et~al.(1991)Yakovlev, Matyugov \& Gubenko}]{Yakovlev1991}
\bibinfo{author}{Yakovlev, O.}, \bibinfo{author}{Matyugov, S.}, \&
  \bibinfo{author}{Gubenko, V.} (\bibinfo{year}{1991}).
\newblock \bibinfo{title}{{Venera-15 and -16 middle atmosphere profiles from
  radio occultations: Polar and near-polar atmosphere of Venus}}.
\newblock {\it \bibinfo{journal}{Icarus}\/},  {\it
  \bibinfo{volume}{94}\/}\bibinfo{issue}{(2)}, \bibinfo{pages}{493--510}.
  \URLprefix
  \url{https://linkinghub.elsevier.com/retrieve/pii/001910359190243M}.
  \DOIprefix\doi{10.1016/0019-1035(91)90243-M}.

\end{thebibliography}

\end{document}